\documentclass[11pt,a4paper]{article} \usepackage{jcappub}
\usepackage{natbib}

\title{Ultra high energy cosmic rays: implications of Auger data for
source spectra and chemical composition}

\author{R. Aloisio$^{1,2}$, V. Berezinsky$^{2,3}$ and P. Blasi$^{1,2}$}

\affiliation{$^{1}$INAF/Osservatorio Astrofisico di Arcetri, Largo
E. Fermi,  5 - 50125 Firenze, Italy\\ $^{2}$Gran Sasso Science
Institute (INFN), viale F. Crispi 7, 67100 L'Aquila,  Italy\\
$^{3}$INFN/Laboratori Nazionali Gran Sasso, ss 17bis km 18+910, 67100
Assergi, Italy\\ }

\emailAdd{aloisio@arcetri.astro.it,berezinsky@lngs.infn.it,blasi@arcetri.astro.it}

\abstract{We use a kinetic-equation approach to describe the
propagation of ultra high energy cosmic ray protons and nuclei and
calculate the expected spectra and mass composition at the Earth for
different assumptions on the source injection spectra and chemical
abundances. When compared with the spectrum, the elongation rate
$X_{max}(E)$ and dispersion $\sigma(X_{max})$ as observed with the
Pierre Auger Observatory, several important consequences can be drawn:
a) the injection spectra of nuclei must be very hard, $\sim
E^{-\gamma}$ with $\gamma\sim 1-1.6$; b) the maximum energy of nuclei
of charge $Z$ in the sources must be $\sim 5Z\times 10^{18}$ eV,
thereby not requiring acceleration to extremely high energies; c) the
fit to the Auger spectrum can be obtained only at the price of adding
an {\it ad hoc} light extragalactic component with a steep injection
spectrum ($\sim E^{-2.7}$). In this sense, at the ankle
($E_{A}\approx 5\times 10^{18}$ eV) all the components are of extragalactic
origin, thereby suggesting that the transition from Galactic to
extragalactic cosmic rays occurs below the ankle. Interestingly, the
additional light extragalactic component postulated above compares
well, in terms of spectrum and normalization, with the one recently
measured by KASCADE-Grande.  }

\begin{document}
\maketitle

\section{Introduction}
\label{sec:intro}

One century after the discovery of Cosmic Rays (CR) a general
description of the basic aspects of the problem has been achieved but
numerous crucial pieces of the puzzle still fail to fall in place. In
the energy range that spans from few GeV/n up to $10^{3}$ TeV/n a self
consistent scenario that accommodates CR composition, propagation and
sources has been developed in the last 30 years, the so-called
standard model of galactic CR (for reviews see
\cite{Ptuskin:2012vs,Blasi:2013p11723} and references therein).

Most CRs with energies below the knee are believed to be produced in
Galactic sources, while their propagation time inside the Galaxy is
energy dependent. The knee reflects a change in the chemical
composition from light to heavy and is well described as a result of
rigidity dependent acceleration and propagation of CRs with gradually
larger mass. The observed CR flux requires a total power of few
percent of the energy released in a supernova (SN) explosion. This
fact is at the very base of Diffusive Shock Acceleration (DSA) models,
that describe CR acceleration in terms of collisionless interactions
of particles with shocks driven by the expanding shell of a supernova
remnant (SNR)
\cite{Blandford:1987p2498,Malkov:2001p172,Caprioli:2010p133}.

The basic version of the SNR paradigm leads to expect that Galactic
CRs should extend to energies of the order of a few $10^{17}$ eV and
that at such energies the chemical composition should be dominated by
iron nuclei. This should be the end of the Galactic CR spectrum.

The origin of ultra high energy CRs (UHECRs) is still enshrouded in
mystery: their sources are unknown, their chemical composition is
subject of much debate and their same definition is somewhat tricky
since it is directly related to the end of Galactic CRs. The
propagation of UHECRs (protons, nuclei and photons) on cosmological
distances has been investigated in detail by many authors and the
underlying physics is well known. The main source of uncertainty in
this problem consists in a poor knowledge of the intergalactic
magnetic fields (if present at all) and the cosmic evolution of the
extragalactic background light (EBL). While the cosmological evolution
of the CMB is known, the evolution with redshift of the EBL should be
inferred from observations at different redshifts through the use of
specific models \cite{Stecker:2005qs,Kneiske:2003tx}. These models are
in good agreement at low redshift ($z<4$), most relevant for the
propagation of UHECR nuclei, while showing significant differences at
high redshift ($z>4$), thereby affecting the production of cosmogenic
neutrinos (see \cite{Allard:2011aa} and references therein). Here we
use the results of \cite{Stecker:2005qs} to model the EBL radiation
field and its cosmological evolution.

 The propagation of UHE nucleons\footnote {Hereafter we will only
consider protons because, as discussed in
\cite{Aloisio:2008pp,Aloisio:2010he,Allard:2011aa}, the decay time of
neutrons is much shorter than all other time scales involved in the
propagation of UHE particles.} is affected only by the interaction
with the CMB radiation field \cite{Aloisio:2010he,Aloisio:2008pp} and
leads to the appearance of two spectral features: the pair-production
dip \cite{Berezinsky:2002nc,Aloisio:2006wv}, which is a rather faint
feature caused by the pair production process ($p+\gamma_{CMB} \to
e^{+} + e^{-} + p$), and a sharp steepening of the spectrum caused by
pion photo-production ($p+\gamma_{CMB} \to \pi + p$) called
Greisen-Zatsepin-Kuzmin (GZK) cutoff
\cite{Greisen:1966jv,Zatsepin:1966jv}. The position of the GZK feature
is roughly defined by the energy where the pair-production and the
photo-pion production rates of energy loss become equal, namely at
$E_{eq}=6.05\times 10^{19}$~eV \cite{Berezinsky:2002nc}, which approximately
corresponds to $E_{GZK} \simeq 50$ EeV

The propagation of UHE nuclei is affected by pair production and
photo-disintegration on the CMB and in general on the EBL. In the
latter process a nucleus with atomic mass number $A$ loses one or more
nucleons because of its interaction with the CMB and the EBL,
$A+\gamma_{CMB,EBL} \to (A-nN) + nN$, being $n$ the number of nucleons
lost by the nucleus \cite{Aloisio:2010he,Aloisio:2008pp}. The
photo-disintegration of nuclei leads to a steepening in the observed
spectrum compared with the injection spectrum. The interplay between
these processes has important consequences in terms of chemical
composition of UHECRs, as discussed below.

From the experimental point of view, measurements of the spectrum and
chemical composition of UHECRs appear to be still not conclusive:
HiRes and Telescope Array (TA) claim that the chemical composition
is compatible with being proton-dominated up to the highest energies.

A somewhat different picture arises from observations carried out with
the Pierre Auger observatory: the published data at the time of writing of this
paper show that the spectrum is not directly understood in terms of a
pure proton composition. The elongation rate and its dispersion, when
used to infer the chemical composition of UHECRs, clearly suggest that
a gradual transition from light to heavy composition takes place
between $10^{18}$ eV and $\sim 5\times 10^{19}$ eV
\cite{Abraham:2010yv} \footnote{A recalibration of the Auger energy
scale is being carried out. Preliminary data presented at the
33$^{rd}$ International Cosmic Ray Conference show a better agreement
with the HiRes and TA spectra at low energies, while a difference in
the GZK region remains. Elongation rate and dispersion are also
affected by the correction in the energy scale but the qualitative
picture is not significantly affected by these new results. The
differences between the two experiments in terms of mass composition
were recently investigated in a critical way by a working group made
of members of the Auger, TA and Yakutsk collaborations
\cite{Barcikowski:2013p3155}.}.

The qualitatively new finding that the mass composition of UHECRs
might be mixed has served as a stimulus to finding propagation models
that can potentially explain the phenomenology of Auger data. Here we
investigate these models and try to address the complex issue of the
transition from Galactic to extragalactic cosmic rays. We show that
the Auger spectrum and mass composition at $E\ge 5\times 10^{18}$ eV
can be fitted at the same time only at the price of requiring very
hard injection spectra for all nuclei ($\sim E^{-\gamma}$ with
$\gamma=1-1.6$) and a maximum rigidity of $\sim 5Z\times 10^{18}
V$. One should appreciate the change of paradigm that this finding
implies: while a decade ago the focus of this field of research was
to find sources and acceleration mechanisms able to energize CR
protons up to energies in excess of $\sim 10^{20}$ eV, present data
require that the UHECR part of the spectrum is made of heavy nuclei
and that protons' maximum energy should not exceed a few
$EeV$. Moreover, rather unusual sources are being implied, in that the
injection spectra are at odds with the predictions of standard
acceleration mechanisms.

However, by accepting these conclusions, it follows that the spectrum
of Auger at energies below $5$ EeV requires an additional component
and we show that it has to be made of light nuclei (protons and helium
nuclei) of extragalactic origin. This implies that at the ankle all
particles are of extragalactic origin, so that the transition from
galactic to extragalactic CRs must occur at somewhat lower energies.

In the energy region $10^{17}-10^{18}$ eV the newly released data of
the KASCADE-Grande (KG) collaboration \cite{Apel:2011mi,Apel:2013ura}
and of the IceTop collaboration \cite{Aartsen:2013p042004}) may help
clarifying the situation.  In
Refs. \cite{Apel:2011mi,Apel:2013ura,Kampert:2013dxa} the presence  of
two separate CR components is discussed: one light (mainly protons and
helium) and the other heavy (mainly iron) with different spectra.  
The results of Icetop \cite{Aartsen:2013p042004} appear to be in
qualitative agreement with those of KG. At $10^{18}$ eV the flux of
light and heavy nuclei appear to have comparable fluxes, in marginal
friction with the results of the three largest UHECR detectors, HiRes,
TA and Auger, that are all compatible with a proton-dominated
composition at EeV energies. This friction should however be taken
with much caution since the absolute fluxes of CRs with different
masses as measured by KG are strongly dependent upon the adopted model
of interactions in the atmosphere, as recently discussed in
\cite{Apel:2014p3320}.

It is however interesting to notice that the light component of KG
fits well in terms of spectrum and absolute flux the one that we
require as additional extragalactic component discussed above and
solely based on fitting the Auger data.

To model the propagation of UHECR we refer to the theoretical
framework based on kinetic equations with a homogeneous distribution
of sources \cite{Aloisio:2010he,Aloisio:2008pp}. The paper is
organized as follows: in \S \ref{sec:propagation} we introduce the
basic theoretical tools and assumptions used in our calculations, in
\S \ref{sec:Auger} we focus on the Auger data and discuss the minimal
requirements needed to achieve a description of the flux and chemical
composition. We also comment on the implications of these models for
the transition from Galactic to extragalactic CRs and compare our
results with the recent data of KASCADE-Grande.
We conclude in \S \ref{sec:conclude}.

\section{Spectra of UHECRs propagating through the CMB and EBL}
\label{sec:propagation}

In this section we briefly summarize the technical aspects of the
calculations carried out below, and we provide references to relevant
work  in which more details can be found.

While propagating through the CMB and the EBL, UHE protons and nuclei
undergo energy losses due to the following processes: expansion of the
universe (adiabatic energy losses), production of $e^++e^-$ pairs,
pion production (for protons) and photo-disintegration (for
nuclei). These processes are treated here in the context of the
continuous energy-loss (CEL) approximation, which {\it a priori} is
not justified for $p+\gamma \to N+\pi$. However, as demonstrated in
\cite{Berezinsky:2002nc}, in the case of a homogeneous distribution of
sources,  the fluctuations in the kinetic equation formalism give a
negligible effect up to energy $10^{21} - 10^{22}$~eV. This is also
true for photo-disintegration  of nuclei, as demonstrated by the good
agreement \cite{Aloisio:2010he} between the kinetic calculations and
different MC schemes
\cite{Aloisio:2012wj,Allard:2005ha,Armengaud:2006fx}. However, the
role of  fluctuations can become important in the case of a discrete
distribution of sources  when the mean distance between sources is
comparable (or larger) with the interaction  length of particles.

The process of photo-disintegration leaves the Lorentz factor of the
nucleus unaltered, and the process can be considered as a 
{\it decay}. The lifetime of a nucleus of atomic mass number $A$ and
Lorentz factor $\Gamma$ at the cosmological time $t$ depends on both
the CMB and EBL background fields and can be written as

\begin{equation}
\frac{1}{\tau_A(\Gamma,t)}=\frac{c}{2\Gamma^2}
\int_{\epsilon_0(A)}^{\infty} d\epsilon_r
\sigma(\epsilon_r,A)\nu(\epsilon_r)\epsilon_r
\int_{\epsilon_r/(2\Gamma)}^{\infty} d\epsilon\;
n(\epsilon,t)/\epsilon^2 ,
\label{eq:tauA}
\end{equation}
where $\epsilon $ is the energy of the target (background) photon in
the laboratory frame and $\epsilon_r$ is the same energy in the rest
frame of the nucleus,  $\epsilon_0(A)$ is the threshold  of the
considered reaction in the rest  frame, $n(\epsilon,t)$ is the density
of the background photons at the cosmological time $t$, given by
$n(\epsilon,t)=n_{\rm CMB}+n_{\rm EBL}$  and $\sigma (\epsilon_r,A)$
is the cross-section of photo-disintegration.  For the EBL we use the
cosmological evolution model of Ref. \cite{Stecker:2005qs}.

In principle the photo-disintegration process involves the emission of
one or more nucleons as described by the multiplicity $\nu(\epsilon)$
in equation (\ref{eq:tauA}). As discussed in
\cite{Aloisio:2008pp,Stecker:1998ib,Puget:1976nz}, the dominant
photo-disintegration channel is the one leading to the emission of one
nucleon ($\nu=1$) that corresponds to the giant dipole resonance in
the photo-disintegration cross section. In \cite{Aloisio:2010he,
Aloisio:2008pp} multinucleon emission ($\nu>1$) was included and found
to be rather small. In this paper we only include the channel with
$\nu=1$.

At large Lorentz factors, photo-disintegration occurs on low energy
CMB photons and the corresponding lifetime of the nucleus
$\tau_A(\Gamma)$ is short, while at small Lorentz factors the lifetime
is dominated by the EBL and $\tau_A(\Gamma)$ is large. The critical
Lorentz factor $\Gamma_c$  where the transition between the two
regimes occurs  is determined by the equality
\begin{equation}
\tau_A^{\rm EBL}(\Gamma_c) = \tau_A^{\rm CMB}(\Gamma_c).
\label{eq:Gamma-crit}
\end{equation}
For $\Gamma > \Gamma_c$ the lifetime sharply decreases and the
spectrum of nuclei becomes steeper, which can be referred to as
photo-disintegration cutoff or Gerasimova-Rozental (GR) cutoff  (see
\cite{GR}).
 
\begin{table}[ht]
\caption{Critical Lorentz factor $\Gamma_c$ and photo-disintegration
cutoff $E_{\rm GR}$. }
\label{table1}
\begin{center}
\begin{tabular}{c|c|c|c|c|c}
\hline nuclei & He$^4$ & N$^{14}$ & Mg$^{24}$ & Ca$^{40}$ &
Fe$^{56}$\\  \hline $\Gamma_c$ & $5\times 10^9$ & $4\times 10^9$ &
$3.5\times 10^9$ & $4\times 10^9$ & $3.2\times 10^9$ \\ \hline $E_{\rm
GR}$ & $2.0\times 10^{19}$ & $5.6\times 10^{19}$ &  $8.4\times
10^{19}$ & $1.6\times 10^{20}$ & $1.8 \times 10^{20}$\\ \hline
\end{tabular}
\end{center}
\end{table}
 
The corresponding critical Lorentz factor $\Gamma_c$ and
photo-disintegration cutoff $E_{GR}=\Gamma_c A m_N$ are shown in Table
\ref{table1} for selected  nuclei.  The critical  Lorentz factor is
approximately the same for all nuclei of interest,  $\Gamma_c \approx
4\times 10^9$, with the largest one for He$^4$ ($\Gamma_c = 5\times
10^9$) and an anomalously low one for Be$^9$ ($6\times 10^8$) and the
next lowest $\Gamma_c = 3.2\times 10^9$ for iron. The constancy of
$\Gamma_c$ is easily understood based on the relation $\Gamma_c
\epsilon_t \sim \epsilon_0$, where $\epsilon_t$ is the typical energy
of a  target photon and $\epsilon_0$ is the binding energy of the
nucleus.

In Tabel \ref{table1} the critical Lorentz factors $\Gamma_c$ and
photo-disintegration cutoffs $E_{GR} = A\Gamma_c m_N$ are listed for
various nuclei.  The approximate independence of $\Gamma_{c}$ on the
type of nucleus has profound implications for the high energy end of
the fluxes of nuclei. Together with $E_{GR}=A\Gamma_c m_N$, the second
energy scale that fixes the flux behavior at the highest energies is
the maximum acceleration energy $E_A^{\max}=ZE_p^{\max}=Z m_N
\Gamma_p^{\max}$, an intrinsic characteristic of the acceleration
process, that hereafter we assume to be rigidity dependent, namely
proportional to the charge $Z$ of the accelerated nucleus.

The photo-disintegration cutoff in the observed spectrum appears if
$E_{GR} < E_A^{\max}$. In the opposite case, $E_{GR} > E_A^{\max}$,
the high energy end of the spectra is not due to photo-disintegration,
because of the lack of particles with $E_A > E_A^{\max}$, and the flux
suppression is due to the acceleration cutoff
$E_A^{\max}=Z\Gamma_p^{\max} m_N$. If one assumes for simplicity that
$A\approx 2 Z$ and $\Gamma_{c}\simeq 4\times 10^{9}$, the condition
$E_{GR}>E_{max}^A$ becomes:
$$ \Gamma_{max}^{p} <  \frac{A}{Z} \Gamma_{c} \sim 8 \times 10^{9}. $$

In this regime, dominated by the maximum acceleration energy, the
photo-disintegration manifests itself as a slow process on the EBL
radiation. This process does not change the Lorentz-factor of nuclei,
which is affected only by the pair production on the CMB radiation
that  is even slower than the photo-disintegration on EBL.

The spectra of nuclei at Earth are calculated by solving analytically
the kinetic transport equations for primary and secondary nuclei and
nucleons.  For protons $p$ and nuclei with mass number $A$ these
equations read:

\begin{equation}
\frac{\partial n_p(\Gamma,t)}{\partial t} - \frac{\partial}{\partial
\Gamma} \left [ b_p(\Gamma,t)n_p(\Gamma,t) \right ] =  Q_p(\Gamma,t)
\label{eq:kin_p}
\end{equation}   
\begin{equation}
\frac{\partial n_{A}(\Gamma,t)}{\partial t} - \frac{\partial}{\partial
\Gamma} \left [ n_{A}(\Gamma,t) b_{A}(\Gamma,t) \right ] +
\frac{n_{A}(\Gamma,t)} {\tau_{A}(\Gamma,t)}  = Q_A (\Gamma,t)
\label{eq:kin_A}
\end{equation}
where $n$ is the equilibrium distribution of particles,
$b=-d\Gamma/dt$ is the rate of decrease of the particle Lorentz
factor,  and  $Q_p$ and $Q_A$ are the production rates per unit
co-moving volume  and time of protons and nuclei, as the sum of those
produced inside the sources and the secondary products of
photo-disintegration. Particles are assumed to be accelerated at
unspecified, homogeneously distributed sources in a rigidity dependent
manner.

For the primary particles we consider three cases of
injection/acceleration. For reasons that will become clear later, we
are especially interested in the case of hard injection, where the
slope of the generation spectrum is $\gamma_g < 2.0$. When normalized
to the energy injected per unit comoving volume and time ${\cal L}_0$,
the CR spectrum reads:

\begin{equation}
Q_{A_0}^{\rm acc}(\Gamma)=\frac{(2-\gamma_g){\cal L}_0}{A_0 m_N}
\frac{1}{\Gamma_{0}^2}\left (\frac{\Gamma}{\Gamma_{0}}  \right
)^{-\gamma_g}~ e^{-\Gamma/\Gamma_{max}},
\label{eq:Qhard}
\end{equation}

with $A_0=1$ in the case of protons. In the special case $\gamma_g=1$
one has:

\begin{equation}
Q_{A_0}^{\rm acc}(\Gamma)= \frac{{\cal L}_0}{A m_N \Gamma_{0}}
\frac{1}{\Gamma}~ e^{-\Gamma/\Gamma_{max}}.
\label{eq:Qlim}
\end{equation}

The spectrum of CRs effectively injected in the intergalactic medium
by a collection of sources can be affected by the convolution of the
spectrum of individual sources and their luminosity and/or maximum
energy achieved. For instance, under realistic assumptions
\cite{Kachelriess:2005xh} the spectra of protons and primary nuclei
can have a broken power-law spectrum. In \cite{Aloisio:2006wv} a
correlation between the source luminosity and the maximum achievable
energy was used to derive an effective injection spectrum with the
following shape:

\begin{equation}  
Q_{A_0}^{\rm acc}(\Gamma)=\frac{{\cal L}_0/(A_0 m_N)}{\ln\Gamma_0 +
1/(\gamma_g-2)}~ q_{\rm gen}(\Gamma)~ e^{-\Gamma/\Gamma_{max}},
\label{eq:broken}
\end{equation}
with
\begin{equation}
q_{\rm gen}(\Gamma)=\left\{
\begin{array}{cc}
1/\Gamma^2 & ~~~~{\rm at}~~ \Gamma \le \Gamma_0\\
\frac{1}{\Gamma_0^2}\left (\frac{\Gamma}{\Gamma_0} \right
)^{-\gamma_g}&~~~~~ {\rm at}~~\Gamma \ge \Gamma_0.
\end{array}\right. 
\label{eq:q_gen}
\end{equation}
Here the low energy slope $\sim 2$ was inspired by the traditional
diffusive shock acceleration canonical result, while the slope
$\gamma_{g}>2$ for Lorentz factor $\Gamma>\Gamma_{0}$ might be  due to
convolution of $\sim E^{-2}$ spectra with different cutoffs from
sources with different luminosities (see for instance
\cite{Kachelriess:2005xh,Aloisio:2006wv}). The value of $\Gamma_{0}$
in (\ref{eq:Qhard}), (\ref{eq:Qlim}) and (\ref{eq:broken}) depends on
the  specific adopted model, but for phenomenological purposes one has
to  assume $\Gamma_0 \sim 10^6 - 10^8$, as it makes the requirements
in  terms of ${\cal L}_0$ less severe.

After acceleration in the sources, nuclei with mass number $A_{0}$
propagate in the intergalactic medium where they can suffer
photo-disintegration, which leads to injection of secondary nuclei
with $A < A_0$ and  secondary protons. As discussed above, the main
channel of photo-disintegration is the extraction of one nucleon
$(A+1)+\gamma \to A + N$. The primary nucleus, the secondary nucleus
and the associated nucleon all have approximately the same Lorentz
factor $\Gamma$, therefore the injection rate of secondary particles
can be written as:
\begin{equation} 
Q^{\rm sec}_A(\Gamma,z)=Q^{\rm
sec}_p(\Gamma,z)=\frac{n_{A+1}(\Gamma,z)}{\tau_{A+1}(\Gamma,z)}.
\label{eq:Q_sec}
\end{equation} 

Equation (\ref{eq:Q_sec}) couples together the transport equations for
nuclei, Eqs. (\ref{eq:kin_A}), that should be solved following the
photo-disintegration chain, namely starting from the solution for the
primary injected nucleus with mass number $A_0$ and then using the
solution to solve the equation for the nuclei with mass number
$A_0-1$, moving downward along the photo-disintegration chain till the
lowest mass secondary nucleus $A=2$ is reached. The equilibrium
distribution for protons $n_p$ is obtained by solving
Eq. (\ref{eq:kin_p}) with both the injection of freshly accelerated
protons and secondary protons produced by the photo-disintegration of
nuclei taken into account:

$$Q_p=Q_p^{\rm acc}+\sum_{A<A_0}Q_{p}^{\rm sec}~.$$

We complete the discussion above by presenting the analytical solution
to equations (\ref{eq:kin_p}) and (\ref{eq:kin_A}), written as a
function of redshift $z$ and particles' Lorentz factor
$\Gamma$. Following \cite{Aloisio:2010he,Aloisio:2008pp} the solution
reads:
\begin{equation}
n_p(\Gamma,z)=\int_{z}^{z_{\rm max}}  \frac{dz'}{(1+z)H(z)}
Q_p^(\Gamma',z') \left (\frac{d\Gamma'}{d\Gamma}\right )_p,
\label{eq:np}
\end{equation}

\begin{equation}
n_{A}(\Gamma,z)=\int_{z}^{z_{max}} \frac{dz'}{(1+z') H(z')}
Q_{A}(\Gamma ',z')\left ( \frac{d\Gamma '}{d\Gamma} \right )_A
e^{-\eta_{A}(\Gamma ',z')},
\label{eq:nA0-solut}
\end{equation}
where $d\Gamma'/d\Gamma$ for protons and nuclei was calculated in
\cite{Berezinsky:1988wi,Aloisio:2010he,Aloisio:2008pp} and the
generation functions $Q_p$ and $Q_A$ for primary and  secondary
particles are described by equations (\ref{eq:Qhard}), (\ref{eq:Qlim})
and (\ref{eq:Q_sec}). The term $e^{-\eta_A}$ in equation
(\ref{eq:nA0-solut}) takes into account the effect of
photo-disintegration of the propagating nucleus $A$ and suppresses the
flux as from our hypothesis of photo-disintegration behaving as a
"decay" process:
\begin{equation} 
\eta_A(\Gamma',z') = \int_z^{z'} \frac{dz''}{(1+z'') H(z'')}
\frac{1}{\tau_A(\Gamma'',z'')} .
\label{eq:etaA}
\end{equation}

We conclude this section summarizing the main parameters and relevant
energy scales of the physics involved in our investigation. The
hypothesis of homogeneously distributed sources with power law
injection fixes the source parameters: emissivity ${\cal L}_0$,
injection power law index $\gamma_g$ and maximum acceleration energy,
assumed to be rigidity dependent, $E^{\max}_A=ZE_{\max}^p=Z m_N
\Gamma_{\max}^p$. The most relevant energy scale in the physics of
propagation of UHE nuclei is the photo-disintegration cut-off
$E_{GR}=Am_N\Gamma_c$. As discussed above, this energy scale affects
the high energy end of the flux of UHE nuclei only in the case of
large enough maximum acceleration energy:
$\Gamma_{\max}^p>(A/Z)\Gamma_c$.

\section{Building models: problems and progress}
\label{sec:Auger}

In this section we build models aimed at describing Auger data on
spectrum and chemical composition of UHECRs.  In this context we adopt
a phenomenological approach in which the  basic source parameters
($\gamma_g$, $E_{max}^p$ and ${\cal L}_0$)  and the relative
abundances of different elements are fitted to Auger data  (both
spectrum and mass composition) with as little as possible {\it a
priori}  theoretical prejudice on what the values of these parameters
should be.

The chemical composition is inferred from the mean value of the depth
of shower maximum $\langle X_{max} \rangle$ and its dispersion (RMS)
$\sigma(X_{max})$. As was first discussed in \cite{Aloisio:2007rc}
(see also \cite{Kampert:2012mx}), the combined analysis of $\langle
X_{max} \rangle$ and $\sigma(X_{max})$ allows one to obtain less model
dependent information on the mass composition of UHECRs. The main
uncertainties in such a procedure are introduced by the
dependence of $\langle X_{max} \rangle$ and its fluctuations on the
interaction models used to describe the shower development. Most
of such models fit low energy accelerator data while providing
somewhat different results when extrapolated to the energies of
relevance for UHECRs (for a review see \cite{Engel:2011zzb} and
references therein).

In our calculations we follow the procedure of Ref.
\cite{Abreu:2013env} where four models of HE interaction were included
to describe the atmospheric shower development, namely EPOS 1.99
\cite{Pierog:2006qv},  Sibyll 2.1 \cite{Ahn:2009wx}, QGSJet 01
\cite{Kalmykov:1997te} and  QGSJet 02 \cite{Ostapchenko:2005nj}, in
order to derive for each  given primary a  simple prescription for
$\langle X_{max} \rangle$ and $\sigma(X_{max})$. The Auger data are
taken from Refs.
\cite{Abreu:2011pj,Salamida:2011zz,Abraham:2010yv}. Shadowed areas are
used to illustrate the uncertainties associated with different models
of shower development in the UHE regime. We stress that below we will
derive only general, most essential, conclusions on different models,
not trying to  obtain a detailed formal fit to the Auger data.

As primary injected particles at the source we consider, additionally
to protons, four classes of primary nuclei, namely Helium, CNO group,
MgAlSi group and Iron. In the computation of the observed flux
generated  by a given primary $A_0$ we include also all secondaries
produced by the  propagation. The flux of all secondaries will be
presented in figures as  gray shadowed regions, while the total flux
generated by a given primary  $A_0$ will be presented as continuous colored lines.

The first important point to make here is that for a homogeneous
distribution of sources,  in order to accommodate a heavy mass
composition at high energy as inferred by Auger, one is forced to
require very hard injection spectra ($\gamma_{g}\leq 1.6$) at the
sources.

\begin{figure}
   \centering \includegraphics[width=0.49\textwidth]{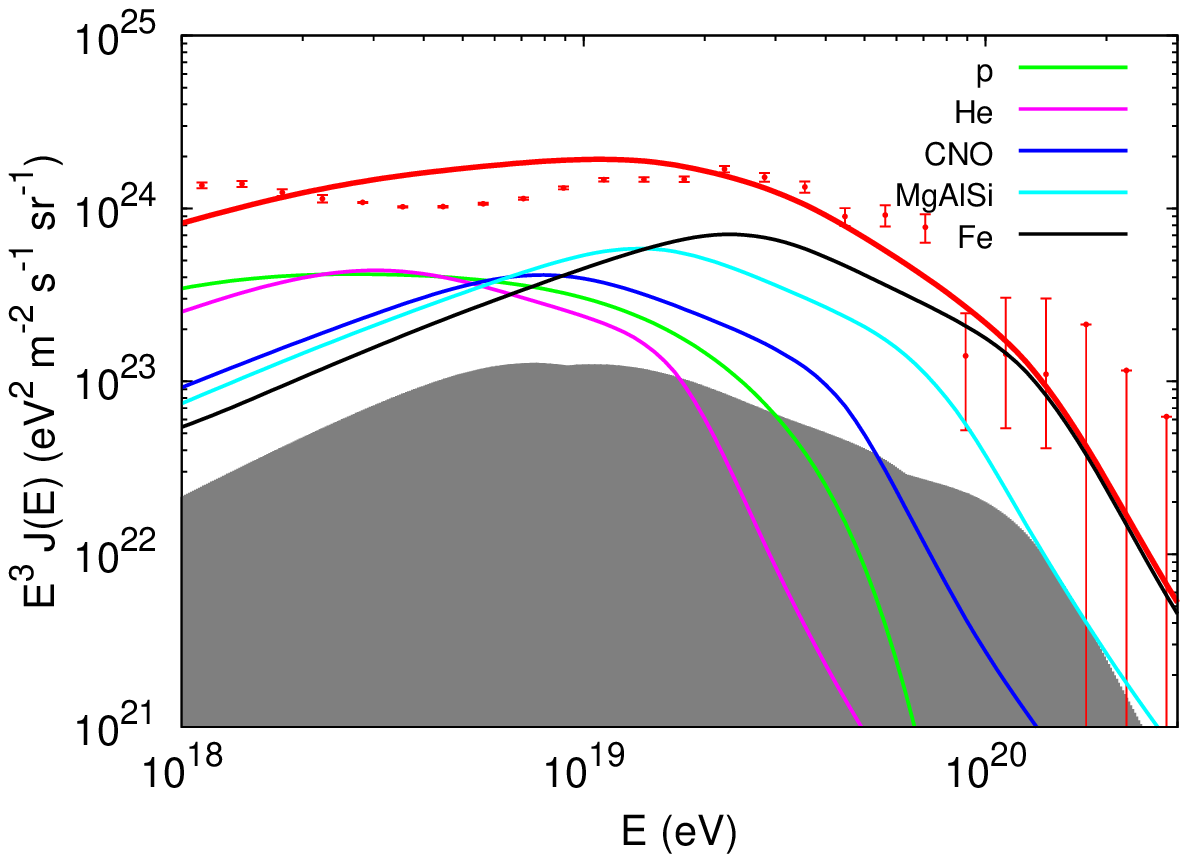}
   \includegraphics[width=0.49\textwidth]{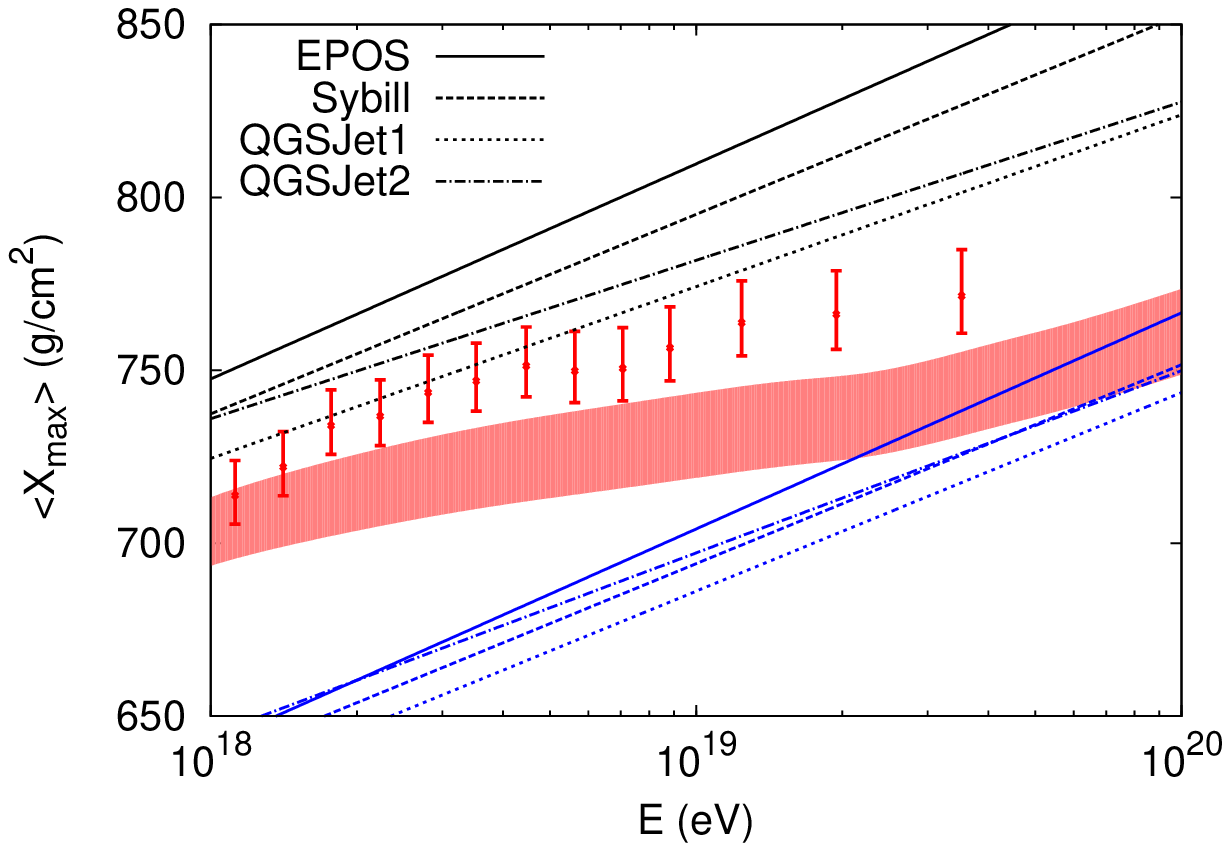}
   \caption{[Left panel] Fluxes of protons and nuclei in the case of
   an injection power law index $\gamma_g=2$ with primary injected
   particles as labelled. Curves with different colors show the sum of
   the flux of primaries with given mass number $A_{0}$ and all
   secondaries produced by the same nuclear species. The thick red
   line shows the total spectrum. The shadowed area shows the flux of
   all secondaries alone. Experimental data are the Auger data on flux
   \cite{Abreu:2011pj,Salamida:2011zz}. [Right panel] Mean value of
   the depth of shower maximum $\langle X_{max} \rangle$ as measured
   by Auger \cite{Abraham:2010yv} and in our calculations (as in the
   left panel).}
   \label{fig:gamma2} 
\end{figure}

In Fig. \ref{fig:gamma2} we illustrate this statement by showing the
case  $\gamma_{g}= 2$. Any steeper injection spectrum makes the
contradiction  even  more severe (see for instance Fig. 5 of
Ref. \cite{Aloisio:2009sj} and Fig. 4c of Ref. \cite{Allard:2011aa} and
references therein).

This difficulty may be described in the following way. The low energy
tail of UHECR spectra reproduces the injection spectrum $n\propto
E^{-\gamma_g}$, as follows from the solution  to the kinetic equation,
Eq. (\ref{eq:nA0-solut}). Therefore, steep injection spectra
($\gamma_g\ge 2$) cause the low energy spectra ($E\gtrsim 10^{18}$ eV)
to be polluted with  heavy nuclei, thereby leading to a disagreement
with the light composition observed by Auger  in the same energy
region. This result is independent of the choice of the maximum
energy.

\begin{figure}
   \centering \includegraphics[width=0.6\textwidth]{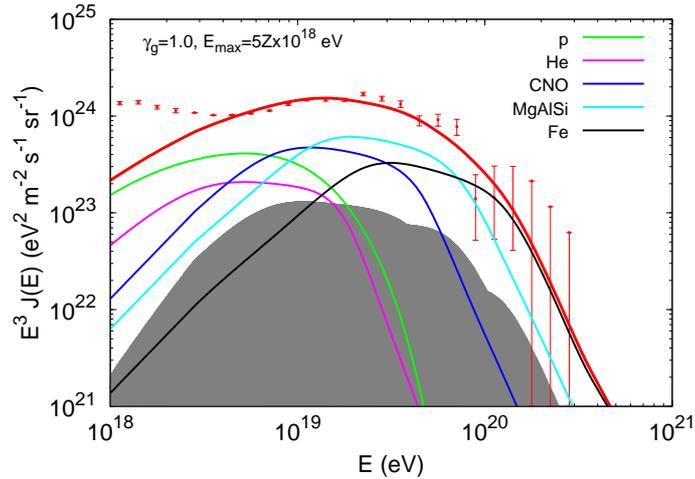}
   \caption{Fluxes of protons and nuclei in the case of an injection
   power law index $\gamma_g=1$ with primary injected particles as
   labelled. Curves with different colors show the sum of the flux of
   primaries with given mass number $A_{0}$ and all secondaries
   produced by the same nuclear species. The shadowed area shows the
   flux of all secondaries alone. Experimental data are the Auger data
   on flux \cite{Abreu:2011pj,Salamida:2011zz}.}
   \label{fig:flux_g1} 
\end{figure}

A hard injection spectrum alleviates this problem.
In Fig. \ref{fig:flux_g1} we plot the spectrum of individual nuclear
species as calculated with the kinetic equation approach  illustrated
above, for $\gamma_g=1$. Again, four classes of primary nuclei
(helium, CNO group, MgAlSi group and Iron) are included (in addition
to protons).

The fit to the Auger data is obtained by using a source emissivity
${\cal L}_0 = 2 \times 10^{44} $ erg/Mpc$^3$/y (above $10^{7}$ GeV/n)
and the following fractions of injection rates relative to proton
injection:

\begin{equation}
Q_{He}^{\rm acc}= 0.2 Q_p^{\rm acc},\;  Q_{CNO}^{\rm acc}= 0.06
Q_p^{acc},\; Q_{MgAlSi}^{\rm acc}= 0.03 Q_p^{acc},\; Q_{Fe}^{\rm acc}
= 0.01 Q_p^{acc}.
\label{eq:Qratios}
\end{equation}

In Fig. \ref{fig:flux_g1} we used the same color codes as in
Fig. \ref{fig:gamma2}, the maximum acceleration energy for protons is
$E_{max}^p=5\times 10^{18}$ eV and for nuclei $E_{max}^A=Z\times
E_{max}^p$.

\begin{figure}
   \centering \includegraphics[width=0.49\textwidth]{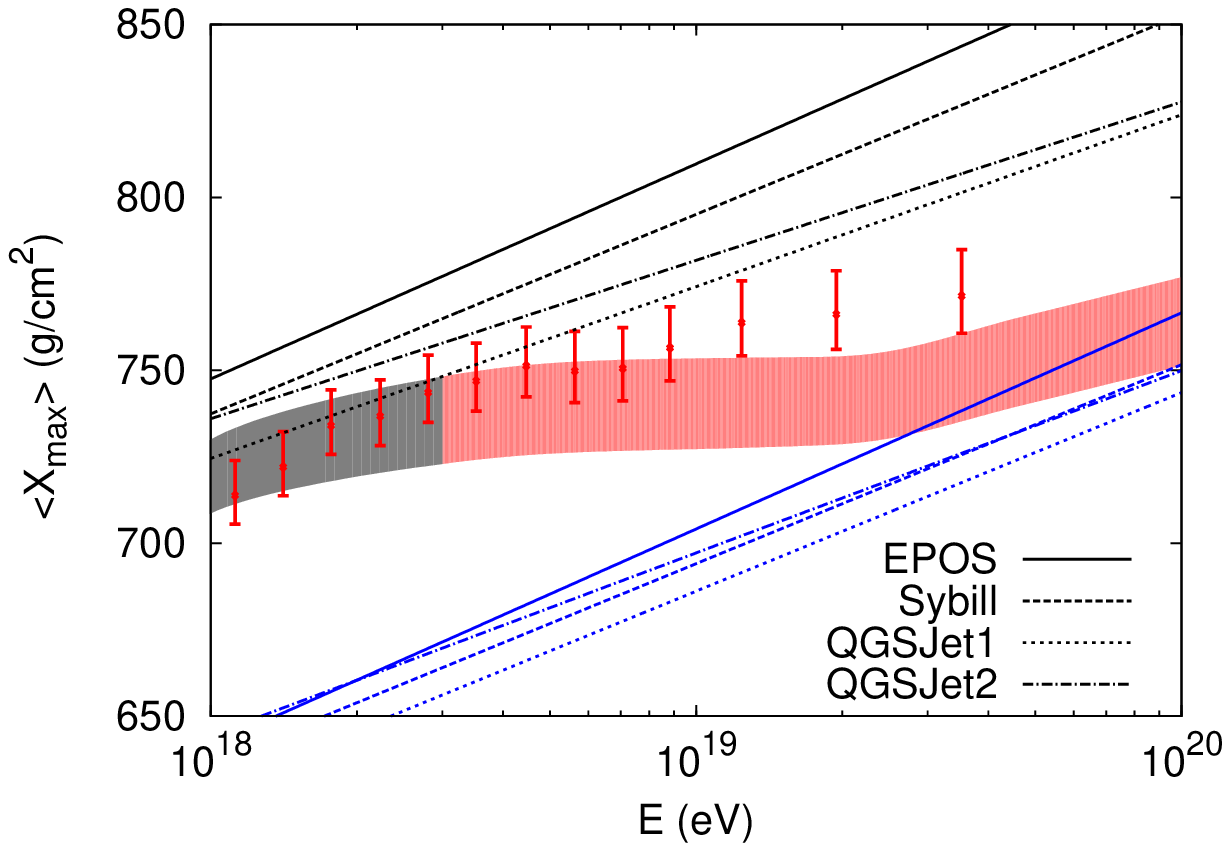}
   \includegraphics[width=0.49\textwidth]{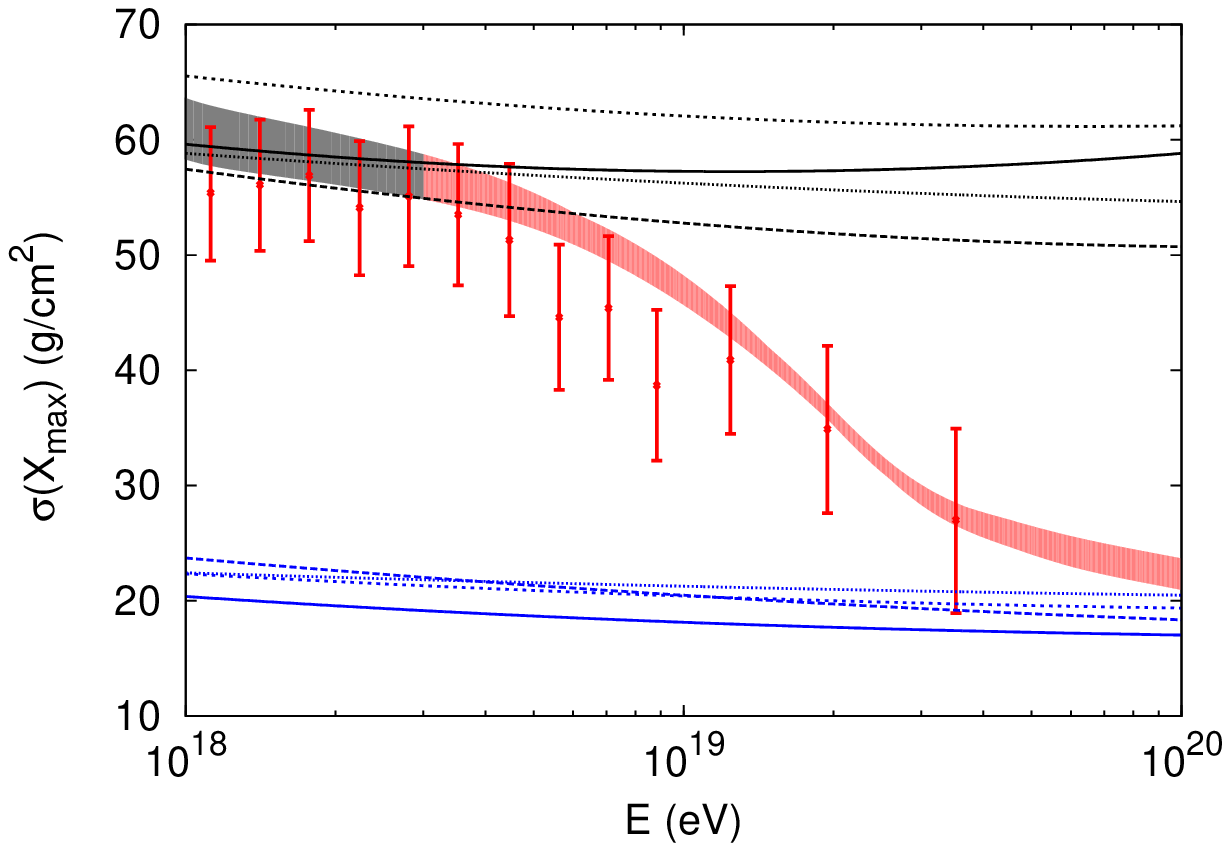}
   \caption{Mean value of the depth of shower maximum $\langle X_{max}
   \rangle$ and its dispersion $\sigma(X_{max})$ as measured by Auger
   \cite{Abraham:2010yv} and in our calculations with the same choice
   of parameters as in figure \ref{fig:flux_g1}. The gray region
   corresponds to the energy range in which the Auger flux is not
   reproduced.}
   \label{fig:chem_g1} 
\end{figure}

The hard injection spectra required to fit the Auger data are
reminiscent of models of the origin of UHECRs associated to
acceleration in rapidly rotating neutron stars
\cite{Blasi:2000xm,Arons:2002yj,Fang:2012rx,Fang:2013cba}, although
hard spectra are a more general characteristic of acceleration
scenarios where regular electric fields are available (e.g. unipolar
induction and reconnection).

The mean depth of shower maximum $\langle X_{max}\rangle$ (left panel)
and its dispersion (right panel) are shown in Fig. \ref{fig:chem_g1}
for the same fluxes as reported in Fig. \ref{fig:flux_g1}. The
combination of hard injection spectrum and low $E_{max}$ allows us to
reach a satisfactory simultaneous agreement with both the spectrum at
$E>5\times 10^{18}$ eV and the chemical composition, though with a
slight tendency toward a too heavy composition at the highest
energies, thereby confirming a result previously presented in Refs.
\cite{Allard:2011aa,Fang:2013cba}.

The fact that a simultaneous fit to the Auger data on the spectrum at
$E>5\times 10^{18}$ eV and the chemical composition requires flat
injection spectra with $\gamma_g\le 1.5-1.6$ also implies that the
spectrum at $E<5\times 10^{18}$ eV cannot be fitted in the same way,
as clearly visible in Fig. \ref{fig:flux_g1}: some kind of new
component in the EeV energy region is required. Hereafter we will
refer to this component as the {\it 'additional EeV component'}. It is
clear that the behavior and chemical composition of this component may
have profound implications for models of the transition from Galactic
to extragalactic CRs.

Below we consider two possibilities to describe the Auger all-particle
CR spectrum and chemical composition, namely the case in which the
additional EeV component has a galactic origin, extending to very high
energies, and the case in which the additional EeV component composed
by protons and helium nuclei has an extragalactic origin.

\subsection{Galactic EeV component} 
\label{subsec:gal}

The basic version of the SNR paradigm suggests that SNRs may
accelerate CRs at the SN blast wave up to  rigidity $R\sim (3-5)
\times 10^{6}$ GV, which leads to iron  nuclei with maximum energy
$7.8\times 10^{16}-1.3\times 10^{17}$ eV.  This range of energies
should also correspond to the end of Galactic CRs. On the other hand
it has been speculated that some rare but more energetic SN events may
give rise to CRs with even larger energies \cite{Ptuskin2010},
although there may be severe theoretical difficulties in understanding
how to achieve such high energies \cite{2013MNRAS.tmp.2026S}.

\begin{figure}
   \centering \includegraphics[width=0.6\textwidth]{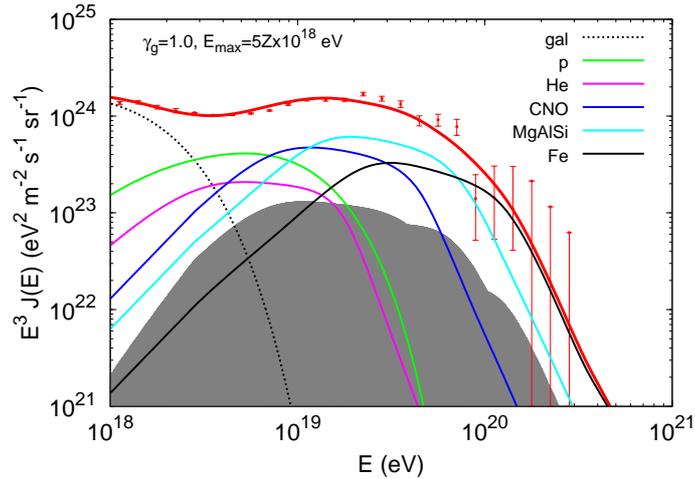}
   \caption{Fluxes of protons and nuclei obtained as in figure
   \ref{fig:flux_g1}. The additional galactic component is plotted as
   dotted black line. Experimental data are the Auger data on flux
   \cite{Abreu:2011pj,Salamida:2011zz}.}
   \label{fig:flux_g1_gal} 
\end{figure}

On the other hand, as discussed above, an additional CR component
appears to be required by the Auger data in the energy range $E <
5\times 10^{18}$ eV,  therefore here we introduce such a component in
the form of a speculative Galactic  CR flux, parametrized as:

\begin{equation}
J_{gal}(E)=J_0 e^{-E/E_\star} \left (\frac{E}{E_\star}\right
)^{-\gamma}
\label{eq:J_gal_new}
\end{equation}
with $E_\star=10^{18}$ eV, $\gamma=2.65$ and $J_0$ chosen in order to
fit the observations. The choice of the power law index $\gamma$ in
equation (\ref{eq:J_gal_new}) comes from the galactic cosmic rays
spectra as computed in \cite{Aloisio:2013tda}. In figure
\ref{fig:flux_g1_gal} we plot the all particle spectrum with the same
choice of the injection parameters used in figure \ref{fig:flux_g1}
and the additional galactic component, Eq. (\ref{eq:J_gal_new}),
plotted as a dotted black curve.

\begin{figure}
   \centering \includegraphics[width=0.49\textwidth]{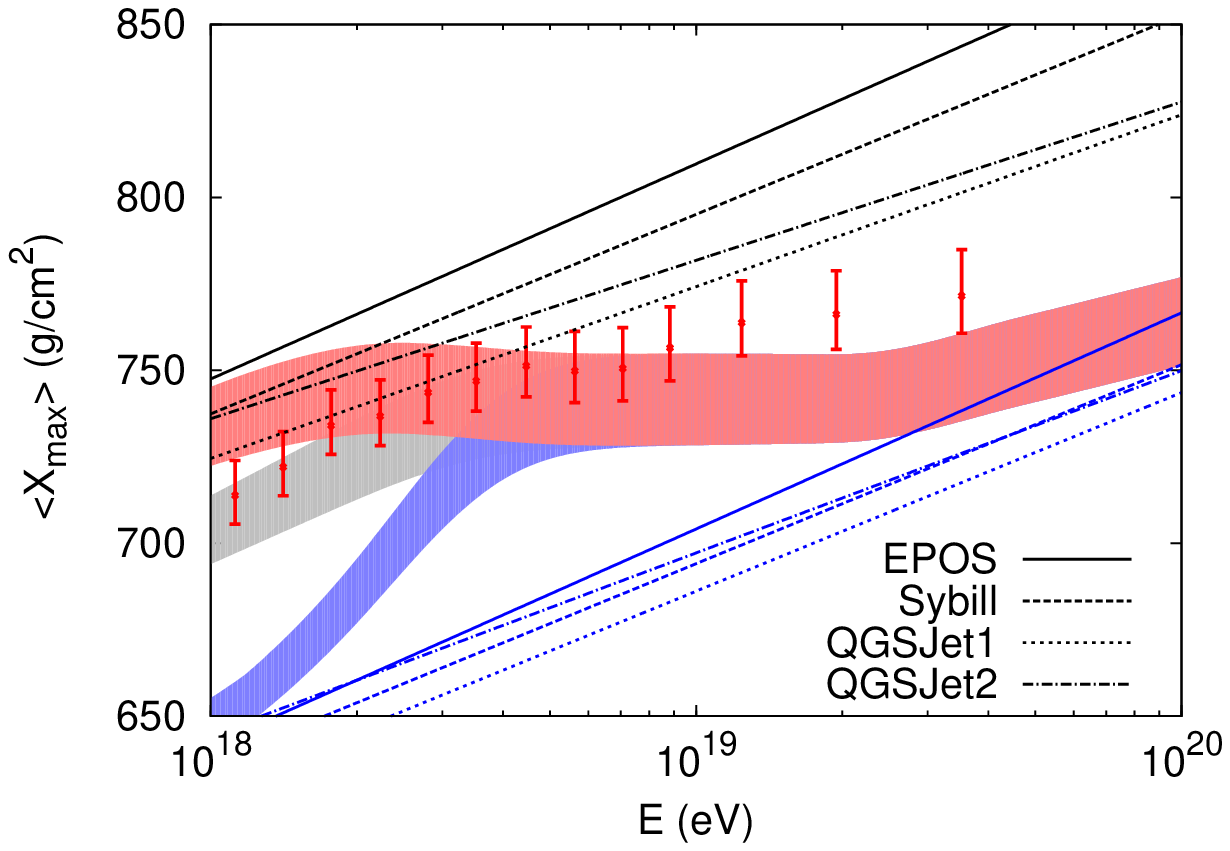}
   \includegraphics[width=0.49\textwidth]{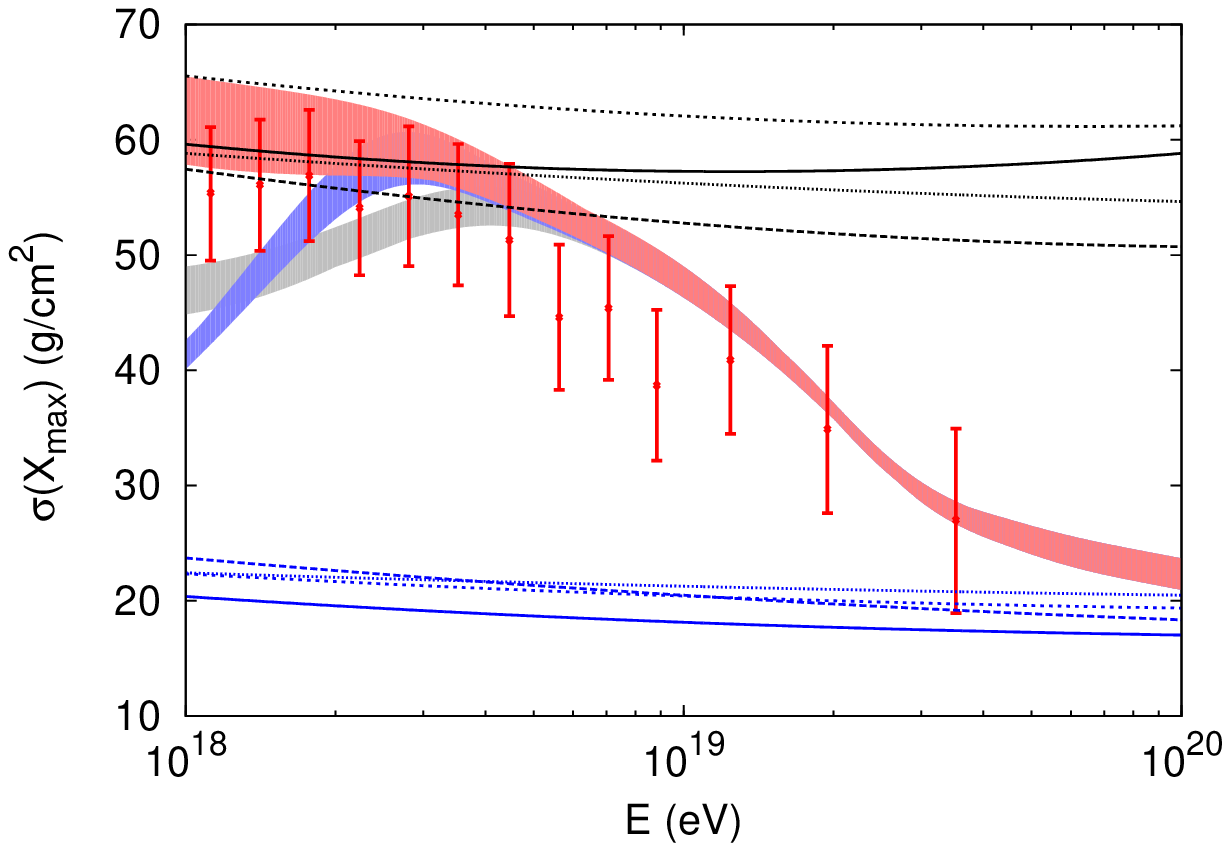}
   \caption{Mean value of the depth of shower maximum $\langle X_{max}
   \rangle$ and its dispersion $\sigma(X_{max})$ as measured by Auger
   \cite{Abraham:2010yv} and in our calculations with the same choice
   of parameters of figure \ref{fig:flux_g1_gal}. The different colors
   of the shadowed regions correspond to the three choices for the
   additional galactic component: protons (red), helium (gray) and
   iron (blu).}
   \label{fig:chem_g1_gal} 
\end{figure}

Given the speculative nature of the Galactic component used here, we
left its chemical composition free, so as to infer it from the
data. More specifically, we consider three possibilities, namely that
the Galactic component consists of protons, helium or iron nuclei. The
corresponding $\langle X_{max}\rangle$ and $\sigma(X_{max})$ are
plotted in Fig. \ref{fig:chem_g1_gal}, where the pure proton
composition is shown as a red shadowed area, the pure helium
composition is shown as a gray shadowed area and, finally, the pure
iron composition of the galactic component is shown as a blue shadowed
area. The latter case would be favored, as far as particle
acceleration is concerned \cite{Ptuskin2010}.

The composition data shown  in  Fig. \ref{fig:chem_g1_gal} strongly
favor a proton composition or a mixture of protons and helium, but
this chemical composition is excluded by the Auger data on anisotropy
\cite{2013ApJ...762L..13P}, since protons and helium in the case
discussed here have a Galactic origin. The anisotropy expected  for a
galactic light component extending up to energies around $10^{18}$ eV
exceeds by more than one order of magnitude the upper limit measured
in \cite{2013ApJ...762L..13P}. The same conclusion was reached by
using a MC simulation of propagation in \cite{2012JCAP...07..031G}.

To satisfy the observations on anisotropy one should assume a pure
iron composition of the additional galactic component, but this case
is  strongly disfavored by $X_{\max}(E)$ as measured by Auger (see the
blue shadowed  area in Fig. \ref{fig:chem_g1_gal}). In fact we tried
to leave  the fraction of Fe nuclei free with respect to protons in
the Galactic additional  component and we found that the RMS at EeV
energies as observed by  Auger can be reproduced only if this fraction
is of order 10\% or smaller.

\subsection{Extragalactic EeV component and KASCADE-Grande light 
component}
\label{subsec:ext}

An additional component of extragalactic light nuclei with a generation spectrum
much steeper than the one used for heavy nuclei can be introduced making use of 
the recent data collected by the KG collaboration, which show the existence 
at sub-EeV energies of a light (p+He) component with a spectral index 
$\gamma=2.79 \pm 0.08$ \cite{Apel:2011mi,Apel:2013ura}. On the other hand, 
the introduction of an extragalactic component of this kind can be done only in 
an artificial way, since there is no theoretical guidance from known acceleration 
models.

To reconcile such steep spectrum with the flat spectrum needed for heavy
nuclei, one has to assume the existence of two classes of sources: one that 
provides only light elements (p+He) with a steep injection and another one
with flat injection that provides also heavy nuclei. 
Here we will only make
the minimal assumption that, in addition to the class of sources that
produce protons and nuclei with flat injection spectrum (as discussed
in the section above),  there is also a second class of sources that
injects light nuclei (protons and helium nuclei)  with a generation
index $\gamma_{g}=2.7$ at high energy
\cite{Berezinsky:2002nc,Aloisio:2006wv,Aloisio:2012ba}.  Both
components are assumed to have a spectral break at Lorentz factor
$\Gamma_{0}=10^{7}$, physically  motivated as in
Refs. \cite{Kachelriess:2005xh} and \cite{Aloisio:2006wv}.

\begin{figure}
   \centering \includegraphics[width=0.49\textwidth]{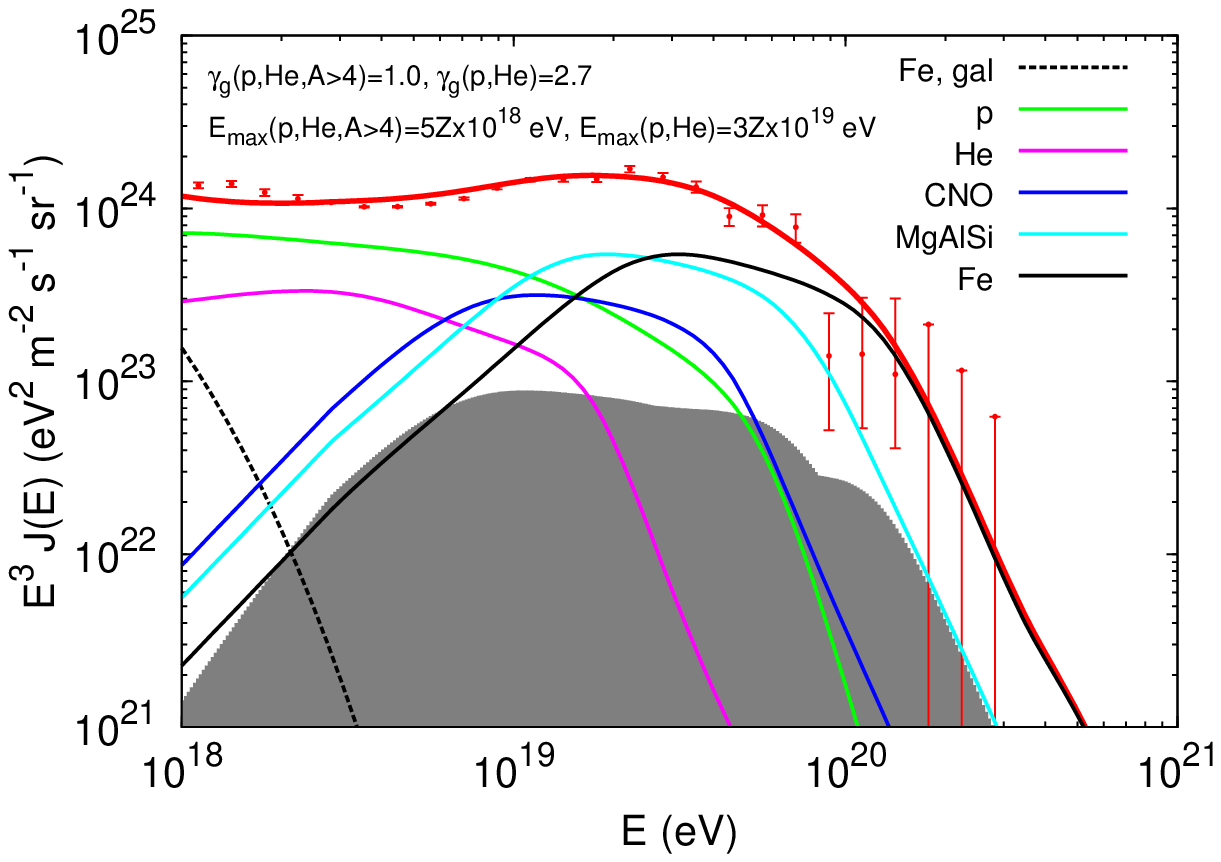}
   \includegraphics[width=0.49\textwidth]{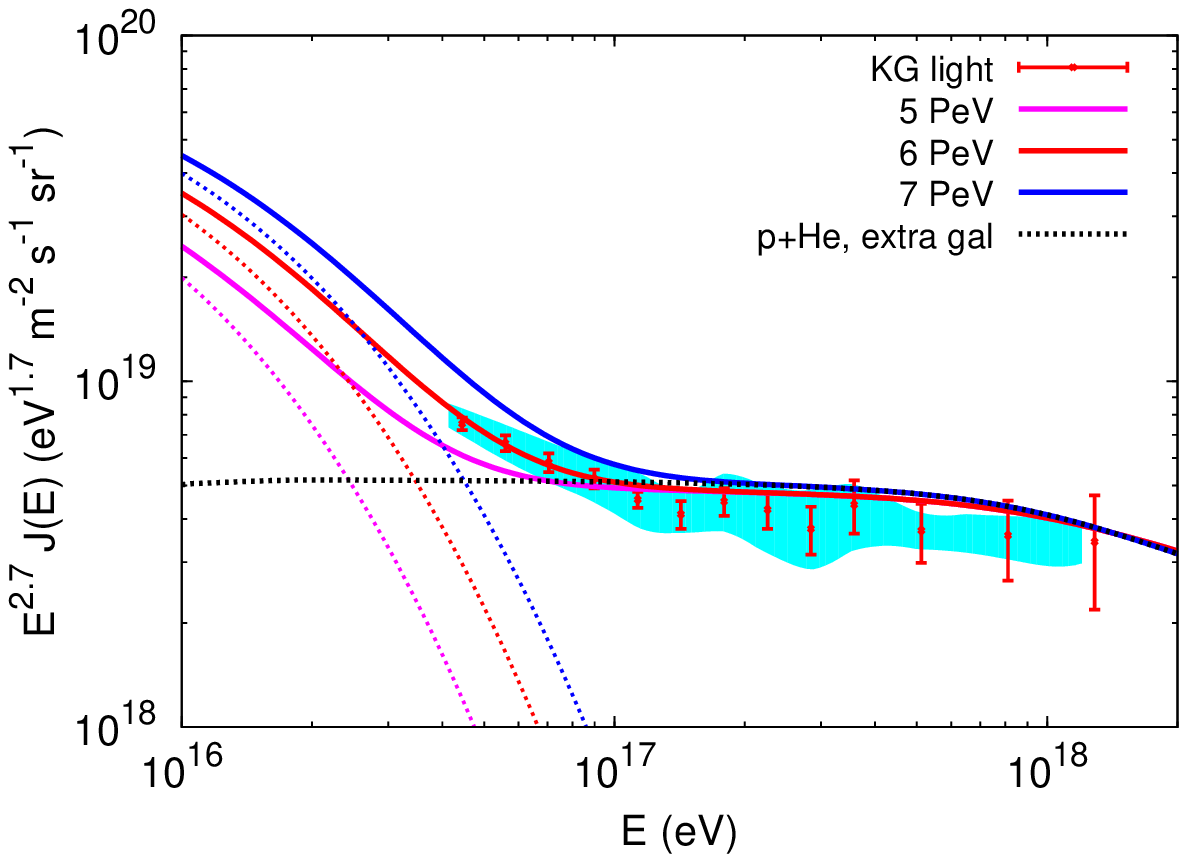}
   \caption{[Left panel] Fluxes of protons and nuclei in the case of
   two populations of extragalactic sources with an injection
   parameters $\gamma_g=2.7, E_{max}^p=3\times 10^{19}$ eV for proton
   and helium and $\gamma_g=1.0, E_{max}^p=5\times 10^{18}$ eV for
   sources providing also  heavier nuclei. Curves with different
   colors show the sum of the flux of primaries with given mass number
   $A_{0}$ and all secondaries produced by the same nuclear
   species. The shadowed area shows the flux of all secondaries alone.
   [Right panel] Kascade grande light component compared with
   extragalactic proton  and helium with $\gamma_g=2.7$ and galactic
   proton and helium fluxes as computed  in \cite{Aloisio:2013tda},
   with three different choices of the maximum acceleration energy  as
   labeled.}
   \label{fig:flux_g1g27}
\end{figure}

\begin{figure}
   \centering \includegraphics[width=0.49\textwidth]{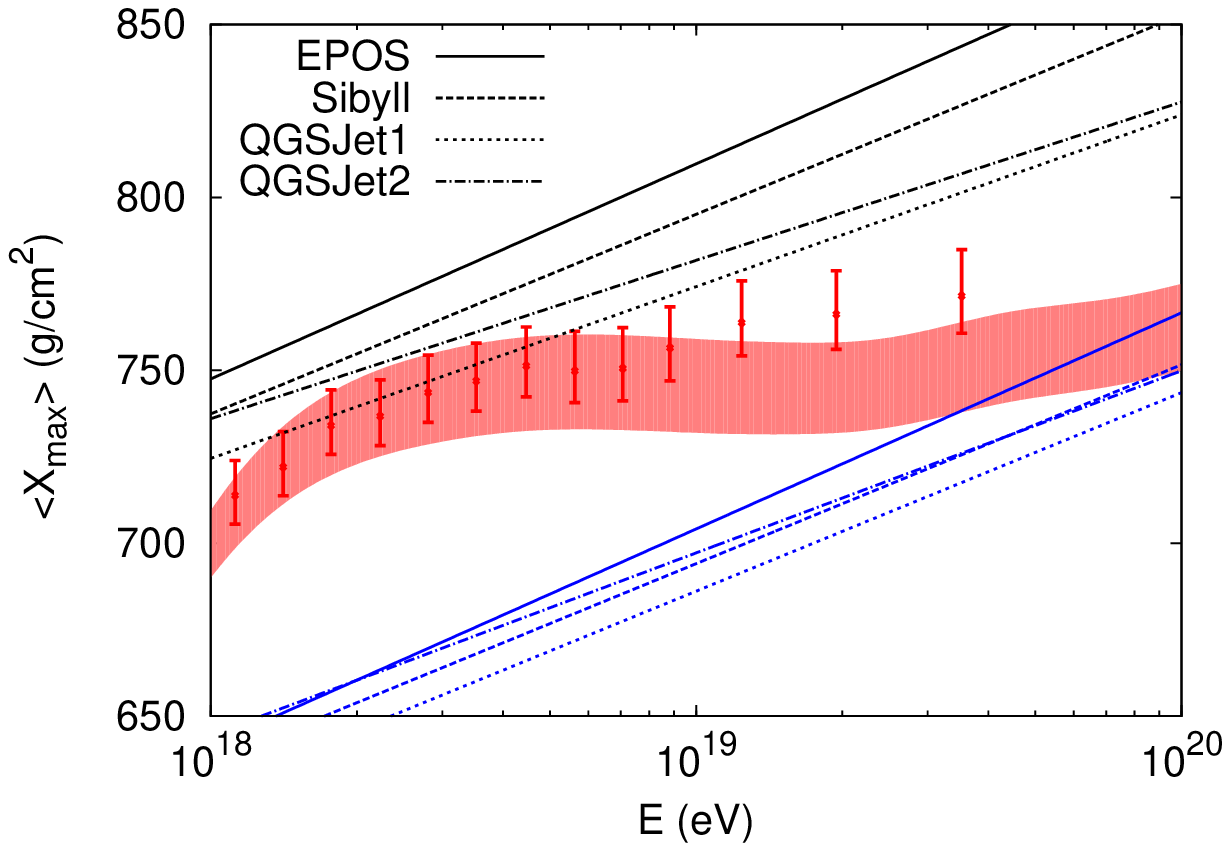}
   \includegraphics[width=0.49\textwidth]{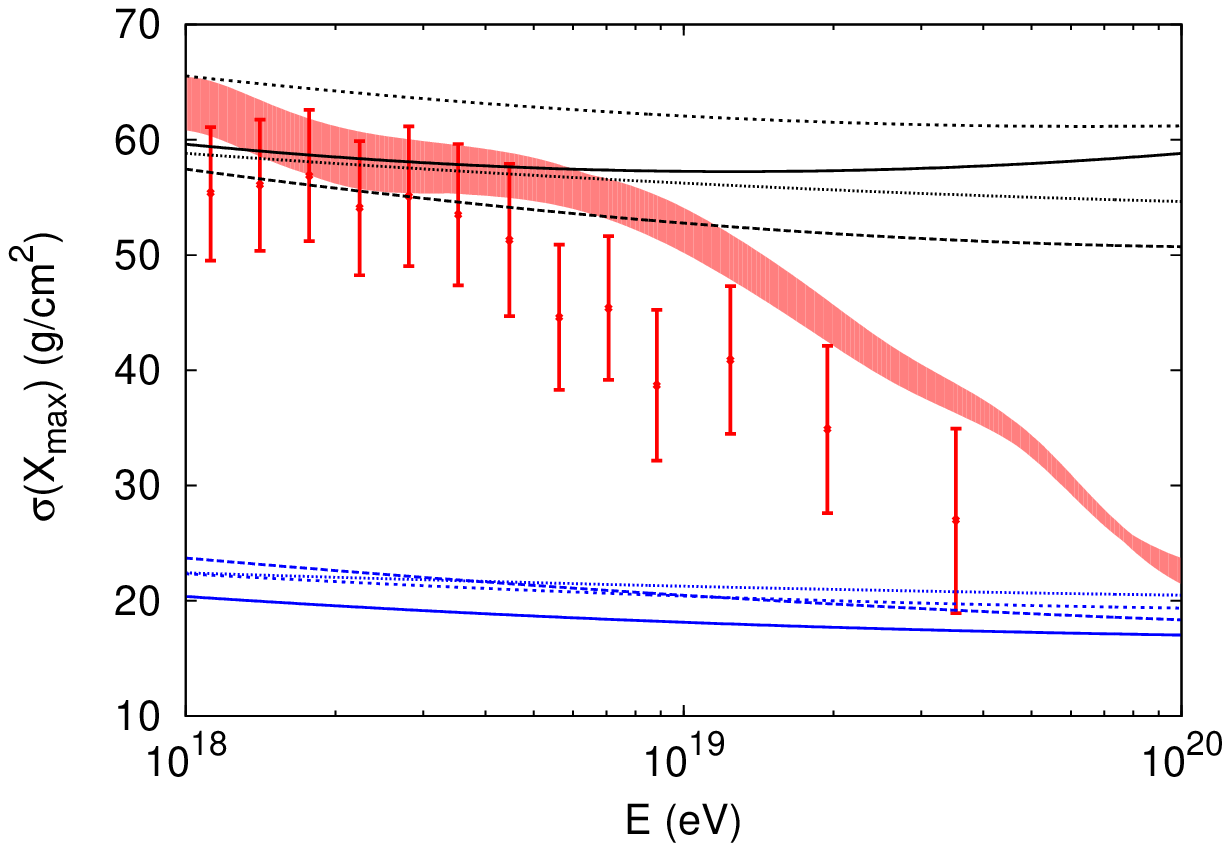}
   \caption{Mean value of the depth of shower maximum $\langle X_{max}
   \rangle$ and its dispersion $\sigma(X_{max})$ as measured by Auger
   \cite{Abraham:2010yv} and in our calculations with the same choice
   of parameters as in figure \ref{fig:flux_g1g27}.}
   \label{fig:chem_g1g27}
\end{figure}

In Figs. \ref{fig:flux_g1g27} (left panel) and \ref{fig:chem_g1g27} we
plot the fluxes and chemical composition ($X_{max}$ and its
dispersion) as obtained with these two classes of sources. We assumed
a source emissivity of the light component (p+He) ${\cal
L}_{0}(p,He)=7\times 10^{49}$ erg/Mpc$^3$/yr (above $10^{7}$ GeV/n),
with a maximum acceleration energy $E_{max}=3Z\times 10^{19}$ eV and
an injection ratio $Q^{\rm acc}_{He}=0.1 Q^{\rm acc}_{p}$. The second
component, contributing p, He, CNO, MgAlSi and Fe, is the same as
discussed in the previous section with a slightly reduced emissivity
${\cal L}_0=1.5 \times 10^{44}$ erg/Mpc$^3$/yr (above $10^{7}$ GeV/n)
and almost the same ratios of injection rates as before (see
Eq. (\ref{eq:Qratios})). 
The total fluxes of p and He are plotted
as thick continuous green and magenta lines respectively, obtained as
the sum of the two contributions to p and He spectra from the two
classes of sources considered. At EeV energies sources providing also
heavy nuclei give a very small contribution to the flux of p and He,
as shown in Fig. \ref{fig:flux_g1}, in this energy range the flux of
light elements (p+He) is contributed mainly by sources with steep
injection. In the left panel of Fig. \ref{fig:flux_g1g27} the end of
the proton spectrum coincides with the maximum energy reached in the
sources, while the spectra of nuclei are ended by photo-disintegration
on the EBL. Together with the extragalactic CR components, in the left panel of 
Fig. \ref{fig:flux_g1g27} we also plot the tail of the galactic (iron dominated) 
CR spectrum (black dotted line) as computed in Ref.  \cite{Aloisio:2013tda} 
(with a maximum energy for galactic protons of 6 PeV, see below).

The fitting to the Auger data on spectrum and mass composition leads
to conclude that at the energy of the ankle, $\sim 5$ EeV, the flux is
dominated by extragalactic CRs, thereby locating the transition from
the Galactic to the extragalactic component in the range $10^{16} -
10^{18}$ eV, with a steep light extragalactic component kicking in
around $\lesssim 10^{18}$ eV.

As anticipated above, it is interesting to notice that a light
CR component has been recently measured by the KASCADE-Grande
collaboration \cite{Apel:2011mi,Apel:2013ura} and attributed to
extragalactic sources. This component is claimed to match the Galactic
light CR spectrum around $10^{17}$ eV. Its spectrum, as measured by
KG, has a slope $\gamma=2.79 \pm 0.08$, compatible with the one
inferred from our calculations based upon fitting the Auger
data. The data points of KG on the light component are shown in the
right panel of Fig. \ref{fig:flux_g1g27}, where the shaded area
illustrates the systematic uncertainties, as estimated by the
collaboration. The rapidly falling dotted lines show the Galactic p+He
spectrum as computed in \cite{Aloisio:2013tda}, with a maximum energy 
of protons of $5$, $6$ and $7$ PeV (see labels). The roughly constant 
black dotted line shows the flux of extragalactic light CRs as calculated 
above, based on the fit to the Auger data. The solid lines indicate the sum 
of the Galactic and extra galactic light components, showing a remarkable 
agreement with the KG data. It is however to be kept in mind that the values 
of the maximum proton energies used here are sizeably higher than the 
standard 3 PeV, and considerably higher than it is easy to obtain from 
theories of CR acceleration at SNR shocks (see for instance
\cite{2013MNRAS.tmp.2026S,Blasi:2014p3323}).

Here we do not want to over-interpret the KG data, since the fluxes
and inferred mass composition are strongly dependent upon the adopted
model for the development of showers in the atmosphere
\cite{Apel:2014p3320}. We only notice that the existence of a light
component that we postulated in fitting Auger data appears naturally
in the KG data.

\section{Discussion and conclusions}
\label{sec:conclude}

In this paper we took the Auger data on the spectrum and chemical
composition of UHECRs at face value and tried to infer as much
physical information as possible.

The evidence that CRs in the energy region $(1-5)\times 10^{18}$ eV
are dominated by light elements may be considered rather solid as it
follows from data on $\langle X_{max}\rangle$ and its dispersion for
the three largest UHECR detectors, Auger \cite{Abraham:2010yv}, TA
\cite{Jui:2011vm,Tsunesada:2011mp} and HiRes
\cite{Abbasi:2002ta,Abbasi:2007sv}. Most of the debate on mass
composition concentrates upon data at energies $\gtrsim 5 \times
10^{18}$ eV.

Here we showed that the spectrum and mass composition of UHECRs as
measured by Auger require a hard injection spectrum at the sources,
with slope  $\gamma_{g}\leq 1.5-1.6$, as also discussed in
Ref. \cite{2014APh....54...48T}. Moreover, we find that the maximum
energy of nuclei with charge Z must be relatively low, $\sim 5\times
10^{18}Z$ eV. One should appreciate the change of paradigm that is
taking place in the aftermath of the Auger data: in the last decade we
moved from a need to accelerate protons up to energies in excess of
$10^{20}$ eV to the requirement to limit the maximum energy of protons
to $\lesssim 5$ EeV.

From the theoretical point of view the hard injection spectrum is
interesting in that it suggests that acceleration mechanisms such as
those that take place in the magnetosphere of rapidly rotating neutron
stars \cite{Fang:2012rx,Fang:2013cba,Arons:2002yj,Blasi:2000xm} or in
the accretion discs around massive black holes
\cite{Blandford:1976mnras} may play a role. Moreover, it is fair to
assume that the environment around a neutron star may be polluted with
heavy nuclei, provided such nuclei can be stripped off the surface of
the star. Propagation of these nuclei in the expanding ejecta of the
parent SN is expected to produce lighter elements because of
spallation and photo-disintegration. As a result, a mixed composition
is expected when CRs escape the source environment
\cite{Fang:2012rx,Fang:2013cba}.

The most disappointing consequence of the hard injection spectra is
that the Auger spectrum can only be fitted for energies $\gtrsim
5\times 10^{18}$ eV, while lower energy CRs require a different
explanation. Filling this gap requires the introduction of an {\it ad
hoc} CR component and we showed here that such component must be 
composed of extragalactic light nuclei (p+He) with an injection spectrum 
with slope $\sim 2.7$. The most straightforward implication of this fact 
is that the transition from Galactic to extragalactic CRs must be taking
place at energies $\lesssim 10^{18}$ eV rather than at the ankle.

One could be tempted to speculate that the necessary extra component
may be provided by Galactic sources, able to accelerate particles to
much higher maximum energies. If these CRs were heavy nuclei, the mass
composition in the energy region $1-5$ EeV would not be in agreement
with the $X_{max}$ and dispersion as observed by Auger (as well as by
HiRes and TA). On the other hand, a light composition of Galactic CRs
at such high energies would conflict with our common wisdom of
particle acceleration in SNRs: even in models involving energetic,
rare SN events \cite{Ptuskin2010} the maximum energy may reach $\sim
10^{18}$ eV but only for iron nuclei and with rather extreme
assumptions. Even if the maximum energy were high enough, the
predicted anisotropy would be wildly in excess of observations
\cite{2013ApJ...762L..13P}. The fair conclusion is to deduce that the
extra component must be made of extragalactic light nuclei (p+He), as
discussed above.

Remarkably this light component has a spectrum and flux which are
compatible with the recently detected flux of light nuclei in the
energy region $10^{16}-10^{18}$ eV by KASCADE-Grande
\cite{Apel:2013ura}. These data show an ankle-like feature at $\sim
10^{17}$ eV, that may be tentatively associated  to the transition to
extragalactic protons.

The disappointing complexity of the viable explanations for the
spectrum and chemical composition of Auger are probably the sign that
the injection spectra needed to fit the data are themselves the result
of a more complex phenomenology. An instance of this could be  the
propagation in extragalactic magnetic fields
\cite{Aloisio:2004fz,Lemoine:2004uw,Aloisio:2009sj,Mollerach:2013dza}
and/or phenomena that  occur inside the sources that may also
potentially affect the spectra of  nuclei injected on cosmological
scales and possibly  preferentially select high energy nuclei.

\section*{Acknowledgments}
We are grateful to the rest of the Arcetri High Energy Astrophysics
Group and to the Auger group of L'Aquila University for continuous
discussions on the subject of UHECR. We also thank K.H. Kampert for
valuable discussions on the KASCADE-Grande data. We also thank the 
anonymous referee that helped us improving the paper. 

\bibliographystyle{JHEP} \bibliography{UHECRbib}

\providecommand{\href}[2]{#2}\begingroup\raggedright\begin{thebibliography}{10}

\bibitem{Ptuskin:2012vs}
V.~Ptuskin, {\it {Propagation of galactic cosmic rays}},  {\em Astropart.Phys.}
  {\bf 39-40} (2012) 44--51.

\bibitem{Blasi:2013p11723}
P.~Blasi, {\it The origin of galactic cosmic rays},  {\em The Astronomy and
  Astrophysics Review} {\bf 21} (Nov, 2013) 70. (c) 2013: Springer-Verlag
  Berlin Heidelberg.

\bibitem{Blandford:1987p2498}
R.~Blandford and D.~Eichler, {\it Particle acceleration at astrophysical
  shocks: A theory of cosmic ray origin},  {\em Physics Reports} {\bf 154}
  (Oct, 1987) 1.

\bibitem{Malkov:2001p172}
M.~A. Malkov and L.~O. Drury, {\it Nonlinear theory of diffusive acceleration
  of particles by shock waves},  {\em Reports on Progress in Physics} {\bf 64}
  (Apr, 2001) 429.

\bibitem{Caprioli:2010p133}
D.~Caprioli, E.~Amato, and P.~Blasi, {\it The contribution of supernova
  remnants to the galactic cosmic ray spectrum},  {\em Astroparticle Physics}
  {\bf 33} (Apr, 2010) 160--168.

\bibitem{Stecker:2005qs}
F.~W. Stecker, M.~Malkan, and S.~Scully, {\it {Intergalactic photon spectra
  from the far ir to the uv lyman limit for $0<z<6$ and the optical depth of
  the universe to high energy gamma-rays}},  {\em Astrophys.J.} {\bf 648}
  (2006) 774--783, [\href{http://xxx.lanl.gov/abs/astro-ph/0510449}{{\tt
  astro-ph/0510449}}].

\bibitem{Kneiske:2003tx}
T.~M. Kneiske, T.~Bretz, K.~Mannheim, and D.~Hartmann, {\it {Implications of
  cosmological gamma-ray absorption. 2. Modification of gamma-ray spectra}},
  {\em Astron.Astrophys.} {\bf 413} (2004) 807--815,
  [\href{http://xxx.lanl.gov/abs/astro-ph/0309141}{{\tt astro-ph/0309141}}].

\bibitem{Allard:2011aa}
D.~Allard, {\it {Extragalactic propagation of ultrahigh energy cosmic-rays}},
  {\em Astropart.Phys.} {\bf 39-40} (2012) 33--43,
  [\href{http://xxx.lanl.gov/abs/1111.3290}{{\tt arXiv:1111.3290}}].

\bibitem{Aloisio:2008pp}
R.~Aloisio, V.~Berezinsky, and S.~Grigorieva, {\it {Analytic calculations of
  the spectra of ultra-high energy cosmic ray nuclei. I. The case of CMB
  radiation}},  {\em Astropart.Phys.} {\bf 41} (2013) 73--93,
  [\href{http://xxx.lanl.gov/abs/0802.4452}{{\tt arXiv:0802.4452}}].

\bibitem{Aloisio:2010he}
R.~Aloisio, V.~Berezinsky, and S.~Grigorieva, {\it {Analytic calculations of
  the spectra of ultra high energy cosmic ray nuclei. II. The general case of
  background radiation}},  {\em Astropart.Phys.} {\bf 41} (2013) 94--107,
  [\href{http://xxx.lanl.gov/abs/1006.2484}{{\tt arXiv:1006.2484}}].

\bibitem{Berezinsky:2002nc}
V.~Berezinsky, A.~Gazizov, and S.~Grigorieva, {\it {On astrophysical solution
  to ultrahigh-energy cosmic rays}},  {\em Phys.Rev.} {\bf D74} (2006) 043005,
  [\href{http://xxx.lanl.gov/abs/hep-ph/0204357}{{\tt hep-ph/0204357}}].

\bibitem{Aloisio:2006wv}
R.~Aloisio, V.~Berezinsky, P.~Blasi, A.~Gazizov, S.~Grigorieva, and B.~Hnatyk,
  {\it {A dip in the UHECR spectrum and the transition from galactic to
  extragalactic cosmic rays}},  {\em Astropart.Phys.} {\bf 27} (2007) 76--91,
  [\href{http://xxx.lanl.gov/abs/astro-ph/0608219}{{\tt astro-ph/0608219}}].

\bibitem{Greisen:1966jv}
K.~Greisen, {\it {End to the cosmic ray spectrum?}},  {\em Phys.Rev.Lett.} {\bf
  16} (1966) 748--750.

\bibitem{Zatsepin:1966jv}
G.~Zatsepin and V.~Kuzmin, {\it {Upper limit of the spectrum of cosmic rays}},
  {\em JETP Lett.} {\bf 4} (1966) 78--80.

\bibitem{Abraham:2010yv}
{\bf Pierre Auger} Collaboration, J.~Abraham {\em et.~al.}, {\it {Measurement
  of the Depth of Maximum of Extensive Air Showers above $10^18$ eV}},  {\em
  Phys.Rev.Lett.} {\bf 104} (2010) 091101,
  [\href{http://xxx.lanl.gov/abs/1002.0699}{{\tt arXiv:1002.0699}}].

\bibitem{Barcikowski:2013p3155}
E.~Barcikowski, J.~Bellido, J.~Belz, Y.~Egorov, S.~Knurenko, V.~de~Souza,
  Y.~Tameda, Y.~Tsunesada, M.~U. for~the Pierre~Auger, T.~Array, and
  Y.~Collaborations, {\it Mass composition working group report at uhecr-2012},
   {\em arXiv} {\bf astro-ph.HE} (Jun, 2013)
  [\href{http://xxx.lanl.gov/abs/1306.4430}{{\tt arXiv:1306.4430}}].

\bibitem{Apel:2011mi}
{\bf KASCADE-Grande Collaboration} Collaboration, W.~Apel {\em et.~al.}, {\it
  {Kneelike structure in the spectrum of the heavy component of cosmic rays
  observed with KASCADE-Grande}},  {\em Phys.Rev.Lett.} {\bf 107} (2011)
  171104, [\href{http://xxx.lanl.gov/abs/1107.5885}{{\tt arXiv:1107.5885}}].

\bibitem{Apel:2013ura}
W.~Apel, J.~Arteaga-Velàzquez, K.~Bekk, M.~Bertaina, J.~Blümer, {\em
  et.~al.}, {\it {Ankle-like Feature in the Energy Spectrum of Light Elements
  of Cosmic Rays Observed with KASCADE-Grande}},  {\em Phys.Rev.} {\bf D87}
  (2013) 081101, [\href{http://xxx.lanl.gov/abs/1304.7114}{{\tt
  arXiv:1304.7114}}].

\bibitem{Aartsen:2013p042004}
M.~G.~A. et~al. (ICETOP~Coll.), {\it Measurement of the cosmic ray energy
  spectrum with icetop-73},  {\em Phys. Rev.} {\bf 88} (2013), no.~D88 042004.

\bibitem{Kampert:2013dxa}
K.-H. {Kampert}, {\it {Ultra-High Energy Cosmic Rays: Results and Prospects}},
  {\em ArXiv e-prints} (May, 2013)
  [\href{http://xxx.lanl.gov/abs/1305.2363}{{\tt arXiv:1305.2363}}].

\bibitem{Apel:2014p3320}
W.~D. Apel, J.~C. Arteaga-Vel{\'a}zquez, K.~Bekk, M.~Bertaina, J.~Bl{\"u}mer,
  H.~Bozdog, I.~M. Brancus, E.~Cantoni, A.~Chiavassa, F.~Cossavella,
  K.~Daumiller, V.~de~Souza, F.~D. Pierro, P.~Doll, R.~Engel, J.~Engler,
  M.~Finger, B.~Fuchs, D.~Fuhrmann, H.~J. Gils, R.~Glasstetter, C.~Grupen,
  A.~Haungs, D.~Heck, J.~R. H{\"o}randel, D.~Huber, T.~Huege, K.-H. Kampert,
  D.~Kang, H.~O. Klages, K.~Link, P.~Łuczak, M.~Ludwig, H.~J. Mathes, H.~J.
  Mayer, M.~Melissas, J.~Milke, B.~Mitrica, C.~Morello, J.~Oehlschl{\"a}ger,
  S.~Ostapchenko, N.~Palmieri, M.~Petcu, T.~Pierog, H.~Rebel, M.~Roth,
  H.~Schieler, S.~Schoo, F.~G. Schr{\"o}der, O.~Sima, G.~Toma, G.~C. Trinchero,
  H.~Ulrich, A.~Weindl, J.~Wochele, M.~Wommer, and J.~Zabierowski, {\it The
  kascade-grande energy spectrum of cosmic rays and the role of hadronic
  interaction models},  {\em ADVANCES IN SPACE RESEARCH} {\bf 53} (May, 2014)
  1456.

\bibitem{Aloisio:2012wj}
R.~Aloisio, D.~Boncioli, A.~Grillo, S.~Petrera, and F.~Salamida, {\it {SimProp:
  a Simulation Code for Ultra High Energy Cosmic Ray Propagation}},  {\em JCAP}
  {\bf 1210} (2012) 007, [\href{http://xxx.lanl.gov/abs/1204.2970}{{\tt
  arXiv:1204.2970}}].

\bibitem{Allard:2005ha}
D.~Allard, E.~Parizot, E.~Khan, S.~Goriely, and A.~Olinto, {\it {UHE nuclei
  propagation and the interpretation of the ankle in the cosmic-ray spectrum}},
   {\em Astron.Astrophys.} {\bf 443} (2005) L29--L32,
  [\href{http://xxx.lanl.gov/abs/astro-ph/0505566}{{\tt astro-ph/0505566}}].

\bibitem{Armengaud:2006fx}
E.~Armengaud, G.~Sigl, T.~Beau, and F.~Miniati, {\it {Crpropa: a numerical tool
  for the propagation of uhe cosmic rays, gamma-rays and neutrinos}},  {\em
  Astropart.Phys.} {\bf 28} (2007) 463--471,
  [\href{http://xxx.lanl.gov/abs/astro-ph/0603675}{{\tt astro-ph/0603675}}].

\bibitem{Stecker:1998ib}
F.~Stecker and M.~Salamon, {\it {Photodisintegration of ultrahigh-energy cosmic
  rays: A New determination}},  {\em Astrophys.J.} {\bf 512} (1999) 521--526,
  [\href{http://xxx.lanl.gov/abs/astro-ph/9808110}{{\tt astro-ph/9808110}}].

\bibitem{Puget:1976nz}
J.~Puget, F.~Stecker, and J.~Bredekamp, {\it {Photonuclear Interactions of
  Ultrahigh-Energy Cosmic Rays and their Astrophysical Consequences}},  {\em
  Astrophys.J.} {\bf 205} (1976) 638--654.

\bibitem{GR}
N.~Gerasimova and I.~Rozental, {\it {Interation of Nuclei and Photons of High
  Energies with a Thermal Radiations in the Universe}},  {\em JETP} {\bf 41}
  (1961) 488.

\bibitem{Kachelriess:2005xh}
M.~Kachelriess and D.~V. Semikoz, {\it {Reconciling the ultra-high energy
  cosmic ray spectrum with Fermi shock acceleration}},  {\em Phys.Lett.} {\bf
  B634} (2006) 143--147, [\href{http://xxx.lanl.gov/abs/astro-ph/0510188}{{\tt
  astro-ph/0510188}}].

\bibitem{Berezinsky:1988wi}
V.~Berezinsky and S.~Grigor'eva, {\it {A Bump in the ultrahigh-energy cosmic
  ray spectrum}},  {\em Astron.Astrophys.} {\bf 199} (1988) 1--12.

\bibitem{Aloisio:2007rc}
R.~Aloisio, V.~Berezinsky, P.~Blasi, and S.~Ostapchenko, {\it {Signatures of
  the transition from Galactic to extragalactic cosmic rays}},  {\em Phys.Rev.}
  {\bf D77} (2008) 025007, [\href{http://xxx.lanl.gov/abs/0706.2834}{{\tt
  arXiv:0706.2834}}].

\bibitem{Kampert:2012mx}
K.-H. Kampert and M.~Unger, {\it {Measurements of the Cosmic Ray Composition
  with Air Shower Experiments}},  {\em Astropart.Phys.} {\bf 35} (2012)
  660--678, [\href{http://xxx.lanl.gov/abs/1201.0018}{{\tt arXiv:1201.0018}}].

\bibitem{Engel:2011zzb}
R.~Engel, D.~Heck, and T.~Pierog, {\it {Extensive air showers and hadronic
  interactions at high energy}},  {\em Ann.Rev.Nucl.Part.Sci.} {\bf 61} (2011)
  467--489.

\bibitem{Abreu:2013env}
{\bf Pierre Auger Collaboration} Collaboration, P.~Abreu {\em et.~al.}, {\it
  {Interpretation of the Depths of Maximum of Extensive Air Showers Measured by
  the Pierre Auger Observatory}},  {\em JCAP} {\bf 1302} (2013) 026,
  [\href{http://xxx.lanl.gov/abs/1301.6637}{{\tt arXiv:1301.6637}}].

\bibitem{Pierog:2006qv}
T.~Pierog and K.~Werner, {\it {Muon Production in Extended Air Shower
  Simulations}},  {\em Phys.Rev.Lett.} {\bf 101} (2008) 171101,
  [\href{http://xxx.lanl.gov/abs/astro-ph/0611311}{{\tt astro-ph/0611311}}].

\bibitem{Ahn:2009wx}
E.-J. Ahn, R.~Engel, T.~K. Gaisser, P.~Lipari, and T.~Stanev, {\it {Cosmic ray
  interaction event generator SIBYLL 2.1}},  {\em Phys.Rev.} {\bf D80} (2009)
  094003, [\href{http://xxx.lanl.gov/abs/0906.4113}{{\tt arXiv:0906.4113}}].

\bibitem{Kalmykov:1997te}
N.~Kalmykov, S.~Ostapchenko, and A.~Pavlov, {\it {Quark-gluon string model and
  EAS simulation problems at ultra-high energies}},  {\em
  Nucl.Phys.Proc.Suppl.} {\bf 52B} (1997) 17--28.

\bibitem{Ostapchenko:2005nj}
S.~Ostapchenko, {\it {Non-linear screening effects in high energy hadronic
  interactions}},  {\em Phys.Rev.} {\bf D74} (2006) 014026,
  [\href{http://xxx.lanl.gov/abs/hep-ph/0505259}{{\tt hep-ph/0505259}}].

\bibitem{Abreu:2011pj}
{\bf Pierre Auger} Collaboration, P.~Abreu {\em et.~al.}, {\it {The Pierre
  Auger Observatory I: The Cosmic Ray Energy Spectrum and Related
  Measurements}},  \href{http://xxx.lanl.gov/abs/1107.4809}{{\tt
  arXiv:1107.4809}}.

\bibitem{Salamida:2011zz}
{\bf Pierre Auger} Collaboration, F.~Salamida, {\it {Update on the measurement
  of the CR energy spectrum above 10**18-eV made using the Pierre Auger
  Observatory}}, .

\bibitem{Aloisio:2009sj}
R.~Aloisio, V.~Berezinsky, and A.~Gazizov, {\it {Ultra High Energy Cosmic Rays:
  The disappointing model}},  {\em Astropart.Phys.} {\bf 34} (2011) 620--626,
  [\href{http://xxx.lanl.gov/abs/0907.5194}{{\tt arXiv:0907.5194}}].

\bibitem{Blasi:2000xm}
P.~Blasi, R.~I. Epstein, and A.~V. Olinto, {\it {Ultrahigh-energy cosmic rays
  from young neutron star winds}},  {\em Astrophys.J.} {\bf 533} (2000) L123,
  [\href{http://xxx.lanl.gov/abs/astro-ph/9912240}{{\tt astro-ph/9912240}}].

\bibitem{Arons:2002yj}
J.~Arons, {\it {Magnetars in the metagalaxy: an origin for ultrahigh-energy
  cosmic rays in the nearby universe}},  {\em Astrophys.J.} {\bf 589} (2003)
  871--892, [\href{http://xxx.lanl.gov/abs/astro-ph/0208444}{{\tt
  astro-ph/0208444}}].

\bibitem{Fang:2012rx}
K.~Fang, K.~Kotera, and A.~V. Olinto, {\it {Newly-born pulsars as sources of
  ultrahigh energy cosmic rays}},  {\em Astrophys.J.} {\bf 750} (2012) 118,
  [\href{http://xxx.lanl.gov/abs/1201.5197}{{\tt arXiv:1201.5197}}].

\bibitem{Fang:2013cba}
K.~Fang, K.~Kotera, and A.~V. Olinto, {\it {Ultrahigh Energy Cosmic Ray Nuclei
  from Extragalactic Pulsars and the effect of their Galactic counterparts}},
  {\em JCAP} {\bf 1303} (2013) 010,
  [\href{http://xxx.lanl.gov/abs/1302.4482}{{\tt arXiv:1302.4482}}].

\bibitem{Ptuskin2010}
V.~Ptuskin, V.~Zirakashvili, and E.-S. Seo, {\it {Spectrum of Galactic Cosmic
  Rays Accelerated in Supernova Remnants}},  {\em ApJ} {\bf 718} (2010) 31.

\bibitem{2013MNRAS.tmp.2026S}
K.~M. {Schure} and A.~R. {Bell}, {\it {Cosmic ray acceleration in young
  supernova remnants}},  {\em MNRAS} (Aug., 2013)
  [\href{http://xxx.lanl.gov/abs/1307.6575}{{\tt arXiv:1307.6575}}].

\bibitem{Aloisio:2013tda}
R.~Aloisio and P.~Blasi, {\it {Propagation of galactic cosmic rays in the
  presence of self-generated turbulence}},  {\em JCAP} {\bf 1307} (2013) 001,
  [\href{http://xxx.lanl.gov/abs/1306.2018}{{\tt arXiv:1306.2018}}].

\bibitem{2013ApJ...762L..13P}
{Pierre Auger Collaboration}, P.~{Abreu}, M.~{Aglietta}, M.~{Ahlers}, E.~J.
  {Ahn}, I.~F.~M. {Albuquerque}, D.~{Allard}, I.~{Allekotte}, J.~{Allen},
  P.~{Allison}, and et~al., {\it {Constraints on the Origin of Cosmic Rays
  above 10$^{18}$ eV from Large-scale Anisotropy Searches in Data of the Pierre
  Auger Observatory}},  {\em ApJ} {\bf 762} (Jan., 2013) L13,
  [\href{http://xxx.lanl.gov/abs/1212.3083}{{\tt arXiv:1212.3083}}].

\bibitem{2012JCAP...07..031G}
G.~{Giacinti}, M.~{Kachelrie{\ss}}, D.~V. {Semikoz}, and G.~{Sigl}, {\it
  {Cosmic ray anisotropy as signature for the transition from galactic to
  extragalactic cosmic rays}},  {\em JCAP} {\bf 7} (July, 2012) 31,
  [\href{http://xxx.lanl.gov/abs/1112.5599}{{\tt arXiv:1112.5599}}].

\bibitem{Aloisio:2012ba}
R.~Aloisio, V.~Berezinsky, and A.~Gazizov, {\it {Transition from galactic to
  extragalactic cosmic rays}},  {\em Astropart.Phys.} {\bf 39-40} (2012)
  129--143, [\href{http://xxx.lanl.gov/abs/1211.0494}{{\tt arXiv:1211.0494}}].

\bibitem{Blasi:2014p3323}
P.~Blasi, {\it Origin of very high and ultra high energy cosmic rays},  {\em
  eprint arXiv} {\bf 1403} (Mar, 2014) 2967.

\bibitem{Jui:2011vm}
{\bf Telescope Array} Collaboration, C.~C. Jui, {\it {Cosmic Ray in the
  Northern Hemisphere: Results from the Telescope Array Experiment}},  {\em
  J.Phys.Conf.Ser.} {\bf 404} (2012) 012037,
  [\href{http://xxx.lanl.gov/abs/1110.0133}{{\tt arXiv:1110.0133}}].

\bibitem{Tsunesada:2011mp}
{\bf Telescope Array} Collaboration, Y.~Tsunesada, {\it {Highlights from
  Telescope Array}},  \href{http://xxx.lanl.gov/abs/1111.2507}{{\tt
  arXiv:1111.2507}}.

\bibitem{Abbasi:2002ta}
{\bf High Resolution Fly's Eye} Collaboration, R.~Abbasi {\em et.~al.}, {\it
  {Measurement of the flux of ultrahigh energy cosmic rays from monocular
  observations by the High Resolution Fly's Eye experiment}},  {\em
  Phys.Rev.Lett.} {\bf 92} (2004) 151101,
  [\href{http://xxx.lanl.gov/abs/astro-ph/0208243}{{\tt astro-ph/0208243}}].

\bibitem{Abbasi:2007sv}
{\bf HiRes} Collaboration, R.~Abbasi {\em et.~al.}, {\it {First observation of
  the Greisen-Zatsepin-Kuzmin suppression}},  {\em Phys.Rev.Lett.} {\bf 100}
  (2008) 101101, [\href{http://xxx.lanl.gov/abs/astro-ph/0703099}{{\tt
  astro-ph/0703099}}].

\bibitem{2014APh....54...48T}
A.~M. {Taylor}, {\it {UHECR composition models}},  {\em Astroparticle Physics}
  {\bf 54} (Feb., 2014) 48--53, [\href{http://xxx.lanl.gov/abs/1401.0199}{{\tt
  arXiv:1401.0199}}].

\bibitem{Blandford:1976mnras}
R.~D. Blandford, {\it {Accretion disc electrodynamics - A model for double
  radio sources}},  {\em MNRAS} {\bf 176} (1976) 465--481.

\bibitem{Aloisio:2004fz}
R.~Aloisio and V.~Berezinsky, {\it {Anti-GZK effect in UHECR diffusive
  propagation}},  {\em Astrophys.J.} {\bf 625} (2005) 249--255,
  [\href{http://xxx.lanl.gov/abs/astro-ph/0412578}{{\tt astro-ph/0412578}}].

\bibitem{Lemoine:2004uw}
M.~Lemoine, {\it {Extra-galactic magnetic fields and the second knee in the
  cosmic-ray spectrum}},  {\em Phys.Rev.} {\bf D71} (2005) 083007,
  [\href{http://xxx.lanl.gov/abs/astro-ph/0411173}{{\tt astro-ph/0411173}}].

\bibitem{Mollerach:2013dza}
S.~Mollerach and E.~Roulet, {\it {Magnetic diffusion effects on the Ultra-High
  Energy Cosmic Ray spectrum and composition}},
  \href{http://xxx.lanl.gov/abs/1305.6519}{{\tt arXiv:1305.6519}}.

\end{thebibliography}\endgroup

\end{document}